\documentclass[prd,twocolumn,showpacs]{revtex4}
\usepackage{graphicx}
\usepackage{float}

\newcommand{\be}{\begin{equation}}
\newcommand{\ee}{\end{equation}}
\newcommand{\ba}{\begin{eqnarray}}
\newcommand{\ea}{\end{eqnarray}}

\def\degr{\hbox{$^\circ$}}

\begin{document}
\draft
\title {Clues for the existence of two $K_1(1270)$ resonances}

\author{L. S. Geng}\thanks{E-mail address: lsgeng@ific.uv.es}
\author{E. Oset}\thanks{E-mail address: oset@ific.uv.es}

\affiliation{Departamento de Fisica Teorica e IFIC, Centro Mixto
Universidad de Valencia-CSIC, Institutos de Investigacion de
Paterna, Apdo 22085, 46071 Valencia, Spain}

\author{L. Roca}\thanks{E-mail address: luisroca@um.es}
\author{J. A. Oller}\thanks{E-mail address: oller@um.es}
\affiliation{\it  Departamento de F\'{\i}sica. Universidad de
Murcia. E-30071 Murcia.  Spain.}
\begin{abstract}
The axial vector meson $K_1(1270)$ was studied within the chiral
unitary approach, where it was shown that it has a two-pole
structure. We reanalyze the high-statistics WA3 experiment $K^-
p\rightarrow K^-\pi^+\pi^- p$ at 63\,GeV, which established the
existence of both $K_1(1270)$ and $K_1(1400)$, and we show that it
clearly favors our two-pole interpretation. We also reanalyze the
traditional K-matrix interpretation of the WA3 data and find that
the good fit of the data obtained there comes from large
cancellations of terms of unclear physical interpretation.

\end{abstract}

\maketitle

\section{Introduction}

Two nonets of spin-parity $1^+$ mesons are expected on the basis of
$L=1$ excitation of $q\bar{q}$ system. According to the particle
data group (PDG)~\cite{Yao:2006px}, they are $b_1(1235)$,
$h_1(1170)$, $h_1(1380)$, $a_1(1260)$, $f_1(1285)$, $f_1(1420)$,
$K_1(1270)$ and $K_1(1400)$. Due to SU(3) breaking, as the mass of
the $s$ quark is larger than those of the $u$ and $d$ quarks, the
$K_1(1270)$ and $K_1(1400)$ are assumed to be a mixture of the SU(3)
eigenstates $K_{1B}$ with $C=-1$ and $K_{1A}$ with $C=+1$. Thus,
they provide a possibility to understand the SU(3) symmetry breaking
in the non-perturbative regime. Particularly important in this
respect is the mixing angle $\theta_K$ between the two SU(3)
eigenstates. In the literature, different approaches have been
adopted to determine its value using various experimental inputs,
but a consensus is not yet reached. Recent BES data even call for
two different values to explain the data, $\theta_K<29\degr$ for
$\psi(2S)$ decay and $\theta_K>48\degr$ for $J/\psi$
decay~\cite{Bai:1999mq}. This issue might become even more
complicated as shown in a recent theoretical study that there might
be two poles for $K_1(1270)$~\cite{Roca:2005nm}--a scenario similar
to that of $\Lambda(1405)$~\cite{Magas:2005vu}. In the present
work, we aim to explore the possible experimental consequence of
such a two-pole structure.

 The $Q$ mesons, i.e. $K_1(1270)$ and $K_1(1400)$ as known today, 
have been observed in $\bar{p}p$ annihilation at
rest~\cite{Armenteros64,Astier:1969dt}, the coherent reaction
$K^+d\rightarrow K^+\pi^+\pi^-d$~\cite{Firestone:1972st}, the baryon
exchange reaction $K^-p\rightarrow
\Xi^-(K\pi\pi)^+$~\cite{Gavillet:1978rj}, the hypercharge exchange
reaction  $\pi^-
p\rightarrow(K\pi\pi)\Lambda$~\cite{Rodeback:1980zt}, the
diffractive productions $K^\pm p\rightarrow K^\pm\pi^+\pi^-
p$~\cite{Brandenburg:1975gv,Daum:1981hb}, and  more lately, in the
decay of $\psi(2S)$ into $K_1(1270)$ and $K_1(1400)$ by BES
collaboration~\cite{Bai:1999mq}, the exclusive decay process
$B\rightarrow J/\psi K_1(1270)$ by Belle
collaboration~\cite{Abe:2001wa}, the mass spectrum and resonant
structure in $\tau^-\rightarrow K^-\pi^+\pi^-\nu_\tau$ decays by
CLEO collaboration~\cite{Asner:2000nx}, and in the decay of
$J/\psi\rightarrow K^*(890)K\pi$~\cite{Bugg:2005ni,Ablikim:2005ni}.
The experimental evidence  can be summarized  as follows: in
diffractive processes one often observes both $K_1(1270)$ and
$K_1(1400)$~\cite{Otter:1976kk,Daum:1981hb,Brandenburg:1975gv}.
However, in non-diffractive processes (such as hypercharge exchange
process~\cite{Rodeback:1980zt} and baryon exchange
process~\cite{Gavillet:1978rj}) one often observes only one
resonance mostly in the $\rho K$
channel~\cite{Rodeback:1980zt,Gavillet:1978rj}. It is interesting to
stress that a two-peak structure has been observed in the $K^*\pi$
invariant mass
spectrum~\cite{Otter:1976kk,Daum:1981hb,Brandenburg:1975gv}. In
Ref.~\cite{Daum:1981hb}, the
 two peaks appear at $\sim1240$\,MeV and $\sim1400$\,MeV.
 While in Ref.~\cite{Brandenburg:1975gv}, the two peaks appear at
 $\sim1200$\,MeV and $\sim1400$\,MeV. The two
 peaks of G. Otter et al.~\cite{Otter:1976kk}, on the other hand, appear at
 $\sim1.27$\,GeV  and $\sim1.37$\,GeV. While such a structure is hardly
 seen in the $K^- p$ reaction at 4.2\,GeV$/c$~\cite{Vergeest:1979jd}.
 Thus, the two-peak structure is clearly related to the reaction
 energy. It seems to be more prominent in high energy $K^\pm p$ reactions than
 in low energy reactions.

 It should be stressed that the most conclusive
 and high-statistics data of $K_1(1270)$ come from the WA3 experiment
 at CERN that accumulated data on the reaction $K^-p\to K^- \pi^+ \pi^- p$
 at 63\,GeV. These data were analyzed by the ACCMOR
 Collaboration~\cite{Daum:1981hb}.
 As will be shown in this paper, the two-peak structure, with a peak
 at lower energy depending drastically on the reaction channel
 investigated,
 can be easily explained in our model with two poles for
 $K_1(1270)$ plus the $K_1(1400)$. With only one pole, as has been noted long time
 ago~\cite{Bowler:1976qe,Daum:1981hb}, there is always a discrepancy for the
 peak positions observed in the $K^*\pi$ and $\rho K$ invariant mass
 distributions. In the present work, we mainly concentrate our study on the
 WA3 data~\cite{Daum:1981hb}. Other data have been carefully studied, but since
 they either have too few events or too much background, no direct contrast of our
 analysis with these data will be presented.

Nowadays, it is generally accepted that QCD is the underlying theory
of strong interactions. Due to the asymptotic freedom, however, its
application at low energies around 1\,GeV is highly problematic.
Therefore, various effective theories have been employed. Chiral
symmetry, related with small $u$, $d$, $s$ masses, provides a
general principle for constructing effective field theory to study
low-energy phenomena. In this respect, Chiral perturbation theory
has been rather successful in studies of low-energy hadron
phenomena~\cite{Weinberg:1978kz,Gasser:1984gg,Meissner:1993ah,Bernard:1995dp,Pich:1995bw,Ecker:1994gg}.
However, pure perturbation theory cannot describe the low-lying
resonances. The breakthrough came with the application of unitary
techniques in the conventional chiral perturbation theory, enabling
one to study higher energy regions hitherto unaccessible, while employing 
chiral Lagrangians. The
unitary extension of chiral perturbation theory, U$\chi$PT, has been
successfully applied to study meson-baryon and meson-meson
interactions. More recently, it has been used to study the lowest
axial vector mesons $b_1(1235)$, $h_1(1170)$, $h_1(1380)$,
$a_1(1260)$, $f_1(1285)$, $K_1(1270)$ and
$K_1(1400)$~\cite{Lutz:2003fm,Roca:2005nm}. Both works generate most
of the low-lying axial vector mesons dynamically but differ in one
thing: In Ref.~\cite{Lutz:2003fm}, the authors claimed to have found
both $K_1(1270)$ and $K_1(1400)$, while in Ref.~\cite{Roca:2005nm},
no signal was found for $K_1(1400)$. In addition, in
Ref.~\cite{Roca:2005nm} the two poles appearing on the second Riemann
sheet were both attributed to $K_1(1270)$ due to the considerations
of pole positions and main decay channels. One should be aware
 that only for low energies U$\chi$PT can be considered model independent 
(either for meson-meson, meson-baryon or baryon-baryon scattering), and
it incorporates the basic symmetries and dynamical features of QCD, among them chiral symmetry 
with its symmetry breaking patterns. At higher energies, the perturbative method of
$\chi$PT is no longer applicable and what U$\chi$PT does is to provide an extrapolation of 
$\chi$PT at higher energies by imposing two restrictions: matching $\chi$PT at low energies
and implementing unitarity in coupled channels in an exact way. These two restrictions give little
freedom to the amplitudes, basically a few subtraction constants in the dispersion relations which are fitted
to experiment.

This paper is organized as follows. In Section II, we briefly
describe the unitary chiral approach. We also explain how we treat
the finite widths of vector mesons. An empirical study is performed
in Section III on the WA3  data. It is demonstrated that the WA3
data can be well explained by our two-pole structure for
$K_1(1270)$. In Section IV, we analyze the K-matrix approach which
has long been used to study the diffractive production of $Q$
mesons. We point out that although this approach can reproduce the
data very well, the results seem to be unstable and not very
meaningful physically.  In Section V, we demonstrate that the most
important channels to describe the WA3 data are the $K^*\pi$ and
$\rho K$ channels. A brief summary is given in Section VI.

\section{Chiral unitary approach}
The detailed formalism has been given in Ref.~\cite{Roca:2005nm}. In
the following, we only provide a brief introduction for the sake of
completeness. In the literature, several unitarization procedures
have been used to obtain a scattering matrix fulfilling exact
unitarity in coupled channels, such as the Inverse Amplitude
Method~\cite{Dobado:1996ps,Oller:1997ng,Oller:1998hw} or the $N/D$
method~\cite{Oller:1998zr}. In this latter work the equivalence with
the Bethe-Salpeter equation used in \cite{Oller:1997ti} was
established.

In the present work we  make use of the Bethe-Salpeter approach,
which leads to the following unitarized amplitude:
\begin{equation}
T=[1+V\hat{G}]^{-1}(-V) \,\vec{\epsilon}\cdot\vec{\epsilon}\,',
\label{bethes}
\end{equation}

\noindent where $\hat{G}=(1+\frac{1}{3}\frac{q^2_l}{M_l^2})G$ is a
diagonal matrix with the $l-$th element, $G_l$, being the two meson
loop function containing a vector and a pseudoscalar meson:
\begin{equation}
G_{l}(\sqrt{s})= i \, \int \frac{d^4 q}{(2 \pi)^4} \,
\frac{1}{(P-q)^2 - M_l^2 + i \epsilon} \,
 \frac{1}{q^2 - m^2_l + i
\epsilon}, \label{loop}
\end{equation}
\noindent with $P$ the total incident momentum, which in the center
of mass frame is $(\sqrt{s},0,0,0)$. In the dimensional
regularization scheme the loop function of Eq.~(\ref{loop}) gives
\begin{widetext}
\begin{eqnarray}
G_{l}(\sqrt{s})&=& \frac{1}{16 \pi^2} \left\{ a(\mu) + \ln
\frac{M_l^2}{\mu^2} + \frac{m_l^2-M_l^2 + s}{2s} \ln
\frac{m_l^2}{M_l^2} \right. \nonumber\\ & & \phantom{\frac{1}{16
\pi^2}} + \frac{q_l}{\sqrt{s}} \left[ \ln(s-(M_l^2-m_l^2)+2
q_l\sqrt{s})+
\ln(s+(M_l^2-m_l^2)+2 q_l\sqrt{s}) \right. \nonumber  \\
& & \left. \phantom{\frac{1}{16 \pi^2} + \frac{q_l}{\sqrt{s}}}
\left. \hspace*{-0.3cm}- \ln(-s+(M_l^2-m_l^2)+2 q_l\sqrt{s})-
\ln(-s-(M_l^2-m_l^2)+2 q_l\sqrt{s}) \right] \right\},\label{propdr}
\end{eqnarray}
\end{widetext}
where $\mu$ is the scale of dimensional regularization. Changes in
the  scale are reabsorbed in the subtraction constant $a(\mu)$, so
that the results  remain scale independent. In Eq.~(\ref{propdr}),
$q_l$ denotes the three-momentum of the vector or pseudoscalar meson
in the center of mass frame.

The tree level amplitudes are calculated using the following
interaction Lagrangian~\cite{Birse:1996hd}:
\begin{equation}
{\cal L}_{I}=-\frac{1}{4}\mathrm{Tr}\left\{\left(\nabla_\mu V_\nu-
\nabla_\nu V_\mu\right) \left(\nabla^\mu V^\nu-\nabla^\nu
V^\mu\right)\right\}, \label{eq:LBirse}
\end{equation}
\noindent where $\mathrm{Tr}$ means SU(3) trace and $\nabla_\mu$ is
the covariant derivative defined as
\begin{equation}
\nabla_\mu V_\nu=\partial_\mu V_\nu+[\Gamma_\mu , V_\nu],
\end{equation}
\noindent where $[,]$ stands for commutator and $\Gamma_\mu$ is the
vector current
\begin{equation}
\Gamma_\mu=\frac{1}{2} (u^\dagger\partial_\mu u +u\partial_\mu
u^\dagger)
\end{equation}
\noindent with
\begin{equation}
u^2=U=e^{i\frac{\sqrt{2}}{f}P}. \label{eq:ufields}
\end{equation}
In the above equations $f$ is the pion decay constant in the chiral
limit and $P$ and $V$ are the SU(3) matrices containing the octet of
pseudoscalar and the nonet of vector mesons respectively:
\begin{equation}
 P \equiv \left(\begin{array}{ccc}
 \frac{1}{\sqrt{2}} \pi^0 + \frac{1}{\sqrt{6}}\eta_8
 & \pi^+ & K^+\\
\pi^-&- \frac{1}{\sqrt{2}} \pi^0 + \frac{1}{\sqrt{6}}\eta_8
& K^0\\
K^-& \bar{K}^0 & -\frac{2}{\sqrt{6}}\eta_8
\end{array}
\right),
\end{equation}
\begin{equation}
 V_\mu \equiv
 \left(\begin{array}{ccc} \frac{1}{\sqrt{2}} \rho^0 +
\frac{1}{\sqrt{2}}\omega
 & \rho^+ & K^{*+}\\
\rho^-& - \frac{1}{\sqrt{2}} \rho^0 + \frac{1}{\sqrt{2}}\omega
& K^{*0}\\
K^{*-}& \bar{K}^{*0} & \phi
\end{array}
\right)_{\mu} . \label{eq:PVmatrices}
\end{equation}

\begin{table}[t]
	\setlength{\tabcolsep}{0.3cm}
	\renewcommand{\arraystretch}{1.2}
\caption{$C_{ij}$ coefficients in isospin basis for the $S=1$,
$I=\frac{1}{2}$ channel.}
\begin{center}
\begin{tabular}{c|ccccc}
\hline\hline
 & $\phi K$ & $\omega K$ & $\rho K$ & $K^* \eta$ & $K^* \pi$ \\
\hline  &&&&&\\[-4.5mm]
$\phi K$   & $0$ & $0$ & $0$ & $-\sqrt{\frac{3}{2}}$ & $-\sqrt{\frac{3}{2}}$   \\
&&&&&\\[-4.5mm]
$\omega K$ & $0$ & $0$ & $0$ & $\frac{\sqrt{3}}{2}$ & $\frac{\sqrt{3}}{2}$   \\
&&&&&\\[-4.5mm]
$\rho K$   & $0$ & $0$ & $-2$ & $-\frac{3}{2}$  & $\frac{1}{2}$    \\
$K^* \eta$ & $-\sqrt{\frac{3}{2}}$ & $\frac{\sqrt{3}}{2}$ & $-\frac{3}{2}$ & $0$ & $0$ \\
&&&&&\\[-4.5mm]
$K^* \pi$  & $-\sqrt{\frac{3}{2}}$ & $\frac{\sqrt{3}}{2}$ & $\frac{1}{2}$ & $0$ & $-2$ \\
\hline\hline
\end{tabular}
\label{tab:Cij5}
\end{center}
\end{table}
The two-vector--two-pseudoscalar amplitudes can be obtained by
expanding the Lagrangian  of Eq.~(\ref{eq:LBirse}) up to two
pseudoscalar meson fields:
\begin{equation}
{\cal L}_{VVPP}=-\frac{1}{4f^2}
\mathrm{Tr}\left([V^{\mu},\partial^{\nu}V_{\mu}]
        [P,\partial_{\nu}P]\right),
\label{eq:L}
\end{equation}
 which would account for the Weinberg-Tomozawa interaction for the
$VP\to VP$ process~\cite{Birse:1996hd,Lutz:2003fm}.  As in
Ref.~\cite{Roca:2005nm} in the pseudoscalar octet we assume
$\eta_8\equiv\eta$. In the vector meson multiplet, ideal
$\omega_1-\omega_8$ mixing is assumed: \be
\phi=\omega_1/\sqrt3-\omega_8\sqrt{2/3},\quad
\omega=\omega_1\sqrt{2/3}+\omega_8/\sqrt{3}. \label{eq:phiw} \ee
 Throughout the work, the following phase convention is used:
  $|\pi^+\rangle=-|1+1\rangle$,
 $|\rho^+\rangle=-|1+1\rangle$, $|K^{-}\rangle=-|1/2-1/2\rangle$ and
 $|K^{*-}\rangle=-|1/2-1/2\rangle$ with the notation $|I I_3\rangle$ to denote isospin states.

From the Lagrangian of Eq.~(\ref{eq:L}) one obtains the $s$-wave
amplitude:
\begin{eqnarray}
V_{ij}(s)&=&-\frac{\epsilon\cdot\epsilon'}{8f^2} C_{ij}
\left[3s-(M^2+m^2+M'^2+m'^2)\right.\nonumber\\
&&\left.\hspace{1.5cm}-\frac{1}{s}(M^2-m^2)(M'^2-m'^2)\right],
\label{eq:Vtree}
\end{eqnarray} where $\epsilon$($\epsilon'$) stands for the polarization
four-vector of the incoming(outgoing) vector meson. The masses
$M(M')$,  $m(m')$ correspond to the initial(final) vector mesons and
initial(final) pseudoscalar mesons respectively, and we use an
averaged value for each isospin multiplet. The indices $i$ and $j$
represent the initial and final $VP$ states respectively. The
$C_{ij}$ coefficients for the $(S,I)=(1,1/2)$ $VP$ channel are
tabulated in Table~\ref{tab:Cij5}.

In Ref.~\cite{Roca:2005nm}, the finite widths of vector mesons are
not taken into account in the dimensional regularization scheme.
They are only considered in the cut-off scheme.  In the present
work, we take into account the finite widths of vector mesons in the
dimensional regularization scheme. The precise analytical structure
of the scheme allows us to calculate the pole positions on the
second Riemann sheet. As we will show below, the main effect of the
widths of vector mesons is to modify the widths of the two poles of
$K_1(1270)$. Since the tree level amplitudes $V$ do not contribute
to the resonances, the widths of the vector mesons contribute
through the momentum $q$ and loop function $G$. 
The appropriate way to implement this, respecting the unitarity implicit in the Bethe-Salpeter
equation, is to substitute in Eq.~(\ref{loop}) the propagator of the unstable particle by its exact propagator,
 incorporating a self-energy that accounts for all the decay channels through its imaginary part.
This is most efficiently done by means of the Lehmann representation which writes the propagator in terms 
of its imaginary part
\begin{equation}\label{lehmann}
	D(s)=\int^\infty_{s_\mathrm{th}} ds_V(-\frac{1}{\pi})\frac{\mathrm{Im}\,D(s_V)}{s-s_V+i\epsilon}
\end{equation}
with $s={q^0}^2-\vec{q}\,^2$ and $s_\mathrm{th}$ the threshold for decay channels, 
and then we take the spectral function for the propagator
$\mathrm{Im}\,D(s_V)$ as
\begin{equation}
\mathrm{Im}\,D(s_V)=\mathrm{Im}\left\{\frac{1}{s_V-M^2_V+iM_V\Gamma_V}\right\}
\end{equation}
where we indulge in the approximation of taking $\Gamma_V$ as constant instead of the explicit function
of $s_V$, which would require detailed study of all the decay channels. This further 
sophisticated step is unnecessary  here, inducing changes far smaller than the uncertainties of the approach from other sources discussed in Ref.~\cite{Roca:2005nm}. By using Eq.~(\ref{lehmann}) the loop function of 
Eq.~(\ref{loop}) now reads 
\begin{widetext}
\begin{equation}
G_{l}(\sqrt{s},M_l,m_l)=\frac{1}{C}\int\limits^{(M_V+2\Gamma_V)^2}_{(M_V-2\Gamma_V)^2}ds_V\times
G(\sqrt{s},\sqrt{s_V},m_l)\times
\left(-\frac{1}{\pi}\right)\mathrm{Im}\left\{\frac{1}{s_V-M^2_V+iM_V\Gamma_V}\right\}
\label{eq15}
\end{equation}
\end{widetext}
with
\begin{equation}
C=\int\limits^{(M_V+2\Gamma_V)^2}_{(M_V-2\Gamma_V)^2}ds_V\times
\left(-\frac{1}{\pi}\right)\mathrm{Im}\left\{\frac{1}{s_V-M^2_V+iM_V\Gamma_V}\right\}
\end{equation}
where a reasonable cut, $M_V\pm 2\Gamma_V$, is done in the $s_V$ integration and the constant $C$ is 
introduced to restore the small loss of renormalization of the Breit-Wigner distribution with this cut.
A similar prescription has been applied to account for the dispersion of the momentum $q$ in some coming
formulas.

In the chiral unitary model that we employed, the free parameters
are the decay constant $f$ and the subtraction constant $a(\mu)$,
which are highly correlated. We have checked that the two-pole
structure remains very robust with respect to a reasonable
readjustment of these parameters. On the other hand, the parameter
values used in Ref.~\cite{Roca:2005nm} produce too low pole
positions for $K_1(1270)$ (see Table~\ref{tab:poles1}).
\begin{table*}[ht]
	\setlength{\tabcolsep}{0.4cm}
	\renewcommand{\arraystretch}{1.2}
\caption{The pole positions and widths of $K_1(1270)$ obtained with
the original parameters of Ref.~\cite{Roca:2005nm}. The constant $f$
is the pion decay constant in the chiral limit, $a(\mu)$ is the
subtraction constant and $\mu$ is the renormalization scale. ``Zero
width'' denotes the results obtained with sharp vector meson masses
while ``Finite width'' denotes the results obtained by taking into
account the finite widths of vector mesons. All energy units are in
MeV.}
\begin{center}
\begin{tabular}{c|c|c|c|c|c|c}\hline\hline
& && \multicolumn{2}{c|}{Zero width}&\multicolumn{2}{c}{Finite
width}\\\cline{4-7} \raisebox{2.3ex}[0pt]{$f$}
&\raisebox{2.3ex}[0pt]{$a(\mu)$}&\raisebox{2.3ex}[0pt]{$\mu$}&1st
pole position&2nd pole position&1st pole position& 2nd pole
position\\\hline
  92 &$-1.85$&900&$(1111-i65)$&$(1216-i4)$&$(1111-i64)$&$(1210-i26)$\\\hline\hline
    \end{tabular}
\label{tab:poles1}
\end{center}
\end{table*}
 Therefore, we have
readjusted these parameters to move the higher pole position closer
to the nominal position of $K_1(1270)$. This can be most
conveniently achieved by increasing $f$.
\begin{figure*}[ht]
\begin{center}
\begin{tabular}{cc}
\includegraphics[scale=0.4]{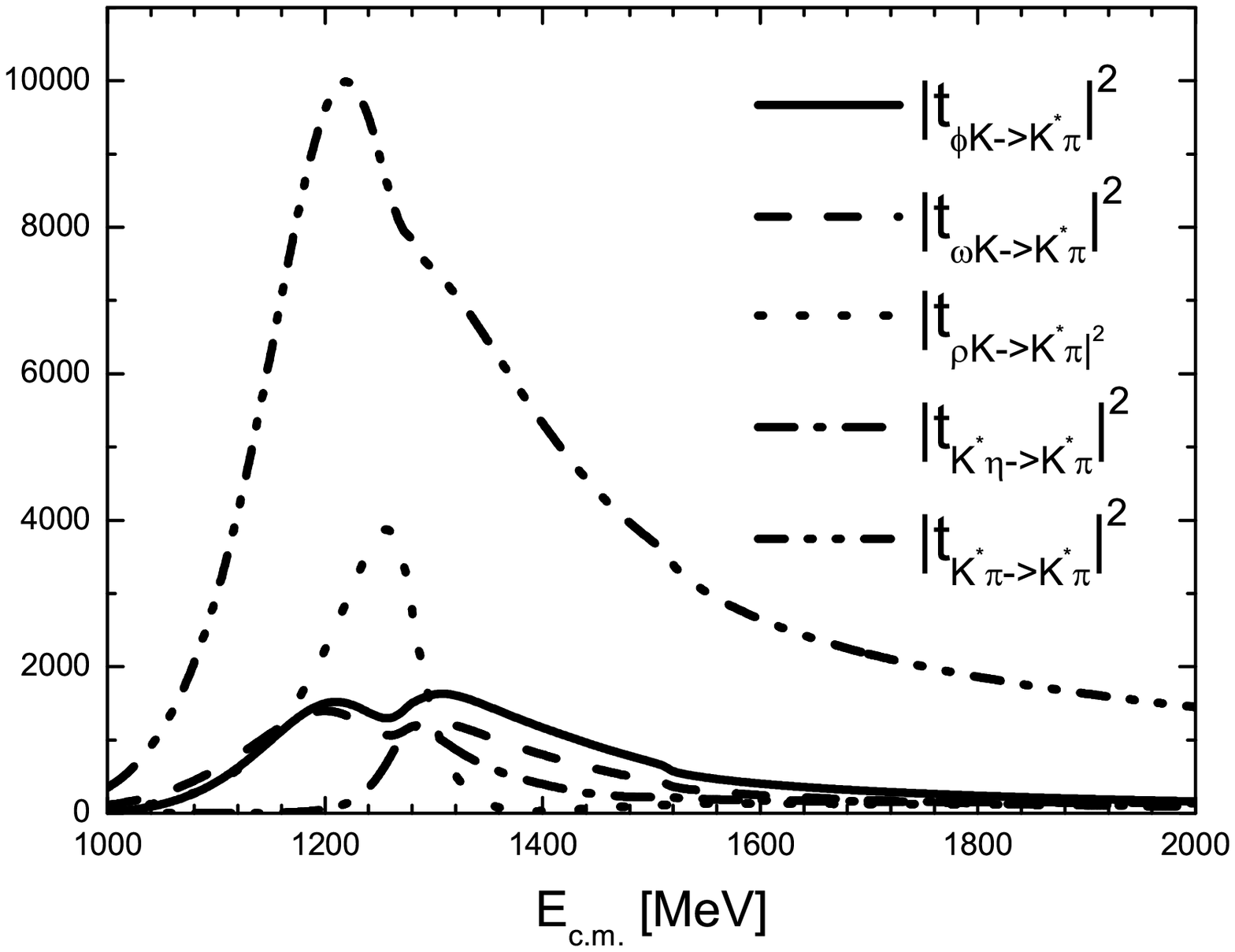} %
&\includegraphics[scale=0.4]{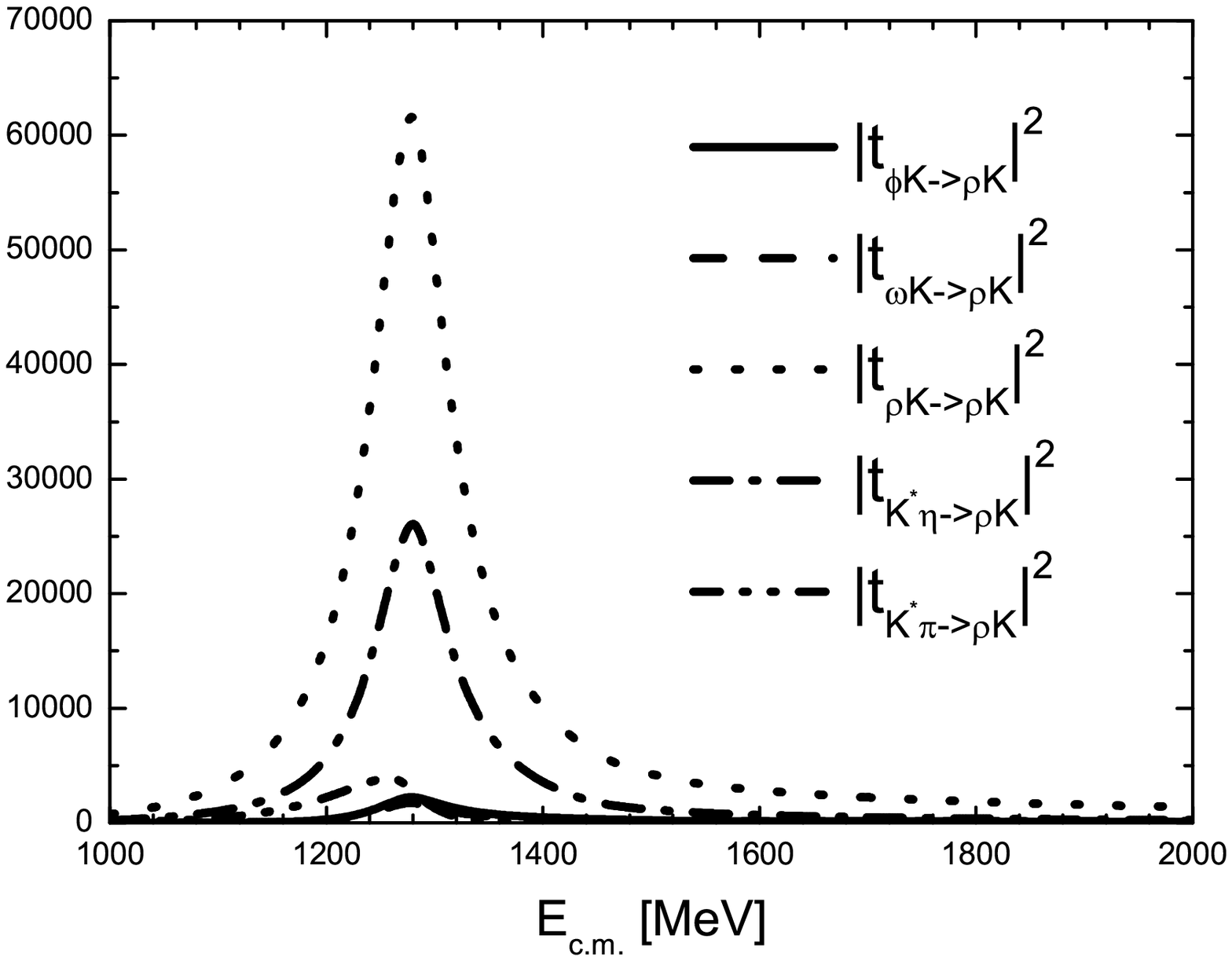}\\
\end{tabular}
\caption{The modulus square of the coupled channel amplitudes in the
$S=1$ and $I=\frac{1}{2}$ channel.}
\label{fig:tVPVP}
\end{center}
\end{figure*}

\begin{figure*}[ht]
\begin{center}
\includegraphics[scale=0.4]{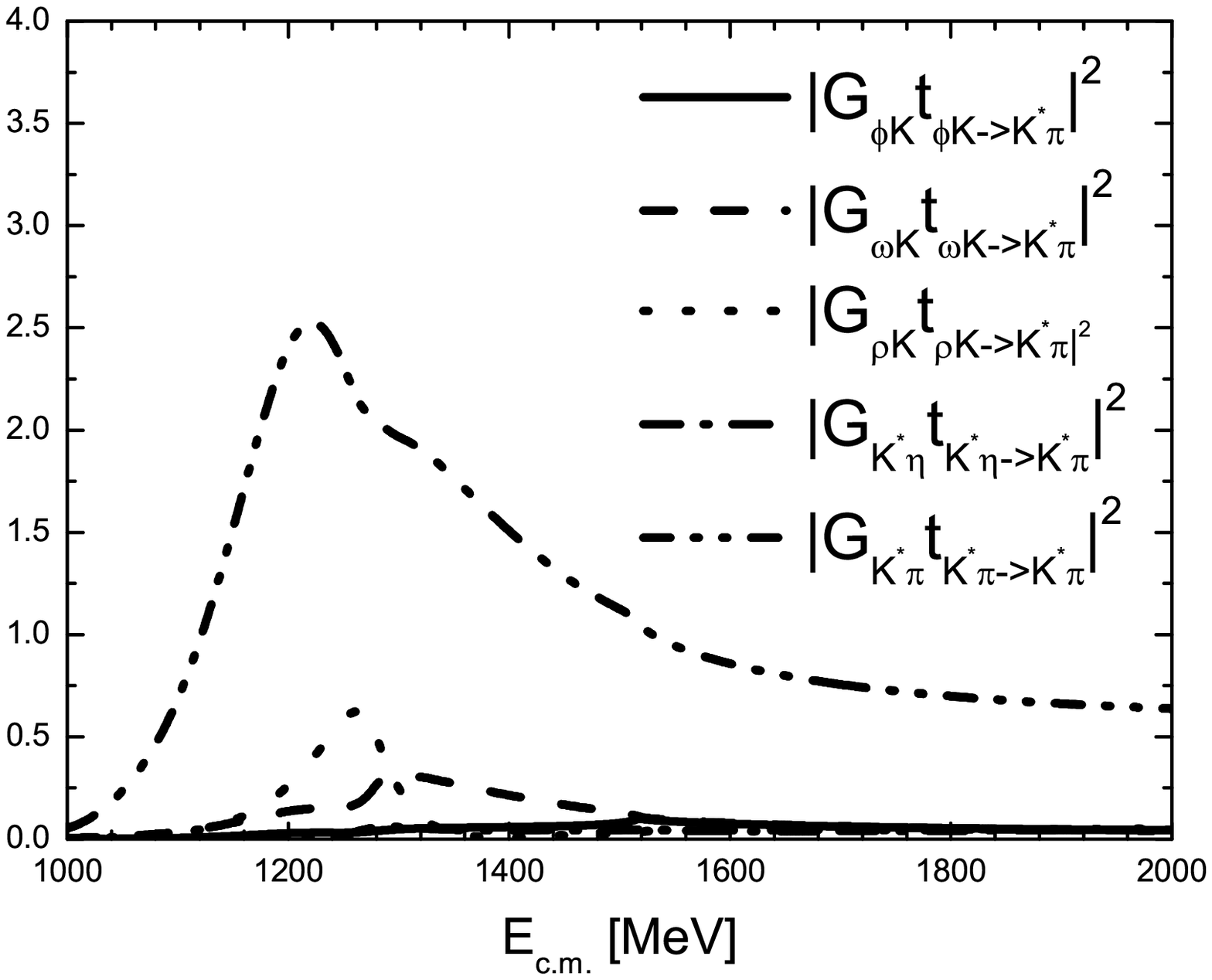}%
\includegraphics[scale=0.4]{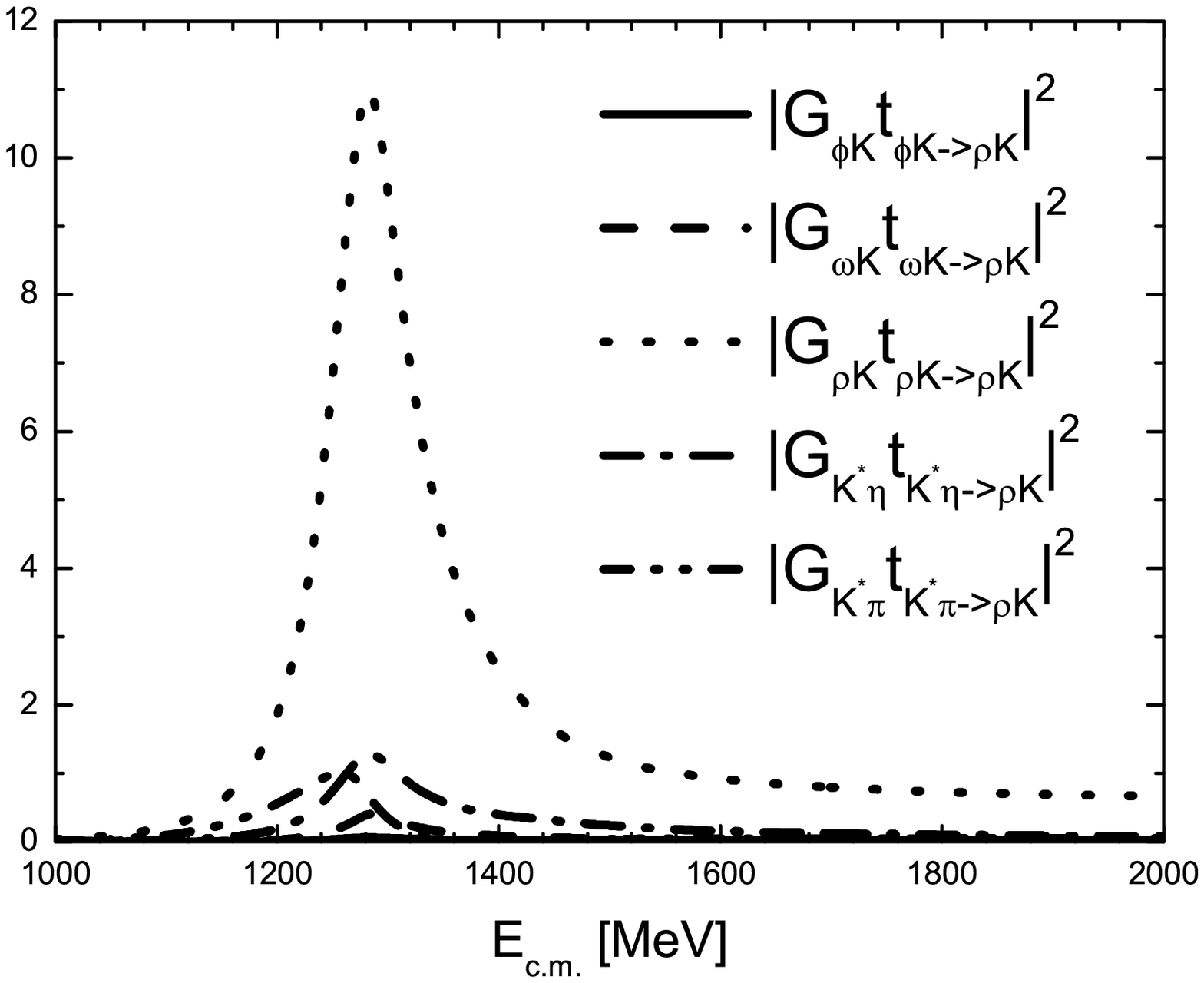}
\caption{The modulus square of the coupled channel amplitudes
multiplied by the corresponding loop functions in the $S=1$ and
$I=\frac{1}{2}$ channel.}
\label{fig:GtVPVP}
\end{center}
\end{figure*}
 Figs.~\ref{fig:tVPVP} and \ref{fig:GtVPVP} show the modulus
square of the $S=1$, $I=\frac{1}{2}$ amplitudes and those multiplied
by the corresponding loop functions obtained with $f=115$\,MeV,
$a(\mu)=-1.85$ and $\mu=900$\,MeV. The pole positions and
corresponding widths obtained with this set of parameters are
tabulated in Table~\ref{tab:poles2}.
\begin{table*}[ht]
	\setlength{\tabcolsep}{0.4cm}
	\renewcommand{\arraystretch}{1.2}
\caption{The same as Table~\ref{tab:poles1},
 but with readjusted parameters.}
\begin{center}
\begin{tabular}{c|c|c|c|c|c|c}\hline\hline
& && \multicolumn{2}{c|}{Zero width}&\multicolumn{2}{c}{Finite
width}\\\cline{4-7} \raisebox{2.3ex}[0pt]{$f$}
&\raisebox{2.3ex}[0pt]{$a(\mu)$}&\raisebox{2.3ex}[0pt]{$\mu$}&1st
pole position&2nd pole position&1st pole position& 2nd pole
position\\\hline
  115&$-1.85$&900&$(1199-i126)$&$(1271-i1)$&$(1195-i123)$&$(1284-i73)$\\\hline\hline
    \end{tabular}
    \label{tab:poles2}
\end{center}
\end{table*}

From Figs.~\ref{fig:tVPVP} and \ref{fig:GtVPVP},
 the two poles are clearly seen:  the higher pole
manifests itself as one relatively narrower resonance around
1.28\,GeV and the lower pole as a broader resonance at
$\sim1.20$\,GeV. Furthermore, these two poles couple to different
channels quite differently. The higher pole is seen mostly in the
$\rho K\rightarrow \rho K$ channel while the lower pole is mostly
seen in the $K^*\pi\rightarrow K^*\pi$ channel. If different
reaction mechanisms favor one or the other channel, they will see
different shapes for the resonance. More importantly, it is to be
noted that not only the two poles couple to different channels with
different strengths, but also they manifest themselves in different
final states. In other words, in the $\rho K$ final states, one
favors a narrower resonance around 1.28\,GeV, while in the $K^*\pi$
final states, one would favor a broader resonance at a smaller
invariant mass.

 As pointed out in Ref.~\cite{Roca:2005nm}, close to a pole, the
$T$ matrix amplitude on the second Riemann sheet can be expressed,
removing the trivial $\epsilon\cdot\epsilon'$ which is also absent
in the other amplitudes, as
\begin{equation}
T_{ij}=\frac{g_i g_j}{s-s_p}
\end{equation}
where $s$ is the energy squared in the center of mass frame and
$\sqrt{s_p}$ the pole position. The numbers $g_i$($g_j$) can be understood as
the effective couplings of the dynamically generated resonance to
channel $i$($j$). They can be calculated from the residues of the
amplitudes at the complex pole positions. The effective couplings
for $\phi K$, $\omega K$, $\rho K$, $K^*\eta$ and $K^*\pi$ are
tabulated in Table~\ref{tab:coup}
 for both the lower pole and the higher pole,
respectively.
\begin{table}[h]
	\setlength{\tabcolsep}{0.3cm}
	\renewcommand{\arraystretch}{1.2}
\caption{Effective couplings of the two poles of $K_1(1270)$ to the
five channels: $\phi K$, $\omega K$, $\rho K$, $K^*\eta$ and
$K^*\pi$. All the units are in MeV.}
\begin{center}
\begin{tabular}{c|cc|cc}
\hline\hline &&&& \\[-4.5mm]
 $\sqrt{s_p}$ & \multicolumn{2}{c|}{$1195-i123$} &\multicolumn{2}{c}{ $1284-i73$ } \\
\cline{2-5}
& $g_i$ & $|g_i|$ & $g_i$ & $|g_i|$ \\
\hline  &&&& \\[-4.5mm]
$\phi K$   &  $2096-i1208$  & $2420$ & $1166-i774$  & $1399$  \\
$\omega K$ &  $-2046+i821$ & $2205$ & $-1051+i620$ & $1220$  \\
$\rho K$   &  $-1671+i1599$& $2313$ & $4804+i395$  & $4821$  \\
$K^* \eta$ &  $72+i197$    & $210$  & $3486-i536$   & $3526$  \\
$K^* \pi$  &  $4747-i2874$ & $5550$ & $769-i1171$   & $1401$  \\
\hline\hline
\end{tabular}
\label{tab:coup}
\end{center}
\end{table}
It is easily seen that the lower pole couples more
dominantly to the $K^*\pi$ channel while the higher pole couples
more strongly to the $\rho K$ channel. We note that these couplings
do not differ qualitatively from those listed in Table~IX of
Ref.~\cite{Roca:2005nm}, although here we have readjusted $f$ from
92\,MeV to 115\,MeV and have taken into account the finite widths of
vector mesons in the dimensional regularization scheme while they
were accounted for in the cutoff method in Ref.~\cite{Roca:2005nm}.

\section{An empirical study of the WA3 data}
As we mentioned in the introduction, the WA3  experiment
$K^-p\rightarrow K^-\pi^+\pi^-p$ at 63\,GeV is one of the most
conclusive and high-statistics experiment on $K_1(1270)$. In this
section, we analyze the WA3  data by constructing production
amplitudes from the $t$ matrix amplitudes obtained in the above
section. The reaction $K^-p\rightarrow K^-\pi^+\pi^- p$ can be
analyzed by the isobar model as $K^-p\rightarrow
(\bar{K}^{*0}\pi^-\,\mbox{or}\,\rho^0 K^-) p\rightarrow
K^-\pi^+\pi^- p$. Therefore, we can construct the following
amplitudes to simulate this process. Assuming $I=\frac{1}{2}$
dominance for $\bar{K}^{*0}\pi^-$ and $\rho^0 K^-$ as suggested by
the experiment we have
\begin{widetext}
\begin{eqnarray}
T_{K^*\pi}\equiv
T_{\bar{K}^{*0}\pi^-}&=&\sqrt{\frac{2}{3}}a+\sqrt{\frac{2}{3}}aG_{K^*\pi}t_{K^*\pi\rightarrow
K^*\pi}+\sqrt{\frac{2}{3}}bG_{\rho
K}t_{\rho K\rightarrow K^*\pi},\nonumber \\
T_{\rho K}\equiv T_{\rho^0
K^-}&=&-\sqrt{\frac{1}{3}}b-\sqrt{\frac{1}{3}}aG_{K^*\pi}t_{K^*\pi\rightarrow
\rho K}-\sqrt{\frac{1}{3}}bG_{\rho K}t_{\rho K\rightarrow \rho K},
\label{eq:Ts}
\end{eqnarray}
\end{widetext} where $t_{ij}$ are the coupled channel amplitudes
obtained in Section II and the Clebsch-Gordan coefficient
$\sqrt{\frac{2}{3}}$($-\sqrt{\frac{1}{3}}$) accounts for projecting
the $I=\frac{1}{2}$ $K^*\pi$ ($\rho K$) state into
$\bar{K}^{*0}\pi^-$($\rho^0 K^-$). The coefficients $a$ and $b$ are
complex couplings. In the most general case, $a$ and $b$ might also
depend on energy. It should be noted that in our chiral unitary
model there are five channels, while in constructing the above
amplitudes, we have only considered two channels $K^*\pi$ and $\rho
K$ due to the following consideration. These two channels are
relatively more important than the other three as can be clearly
seen from Fig.~\ref{fig:GtVPVP}. In the $\rho K$ channel,
$|G_{K^*\eta}t_{K^*\eta\rightarrow \rho K}|^2$ is of similar
magnitude as that of $|G_{K^*\pi}t_{K^*\pi\rightarrow \rho K}|^2$,
but both are much smaller than $|G_{\rho K}t_{\rho K\rightarrow \rho
K}|^2$. Therefore, we expect in the $\rho K$ channel, one will
almost always observe a narrow resonance at $\sim 1280$\,MeV.
Similar argument can be made about the $K^*\pi$ channel. This
consideration allows us to reduce the number  of free parameters,
which would otherwise increase linearly with the number of channels
included.

To contrast our model with data, it is necessary for us to take into
account the existence of $K_1(1400)$, which is not dynamically
generated in our approach. Therefore, we add to the amplitudes in
Eq.~(\ref{eq:Ts}) an explicit contribution of $K_1(1400)$
 \begin{eqnarray}
 T_{K^*\pi}&\rightarrow&
 T_{K^*\pi}+\frac{g_{K^*\pi}}{s-M^2+iM\Gamma(s)},\nonumber\\
  T_{\rho K}&\rightarrow&
 T_{\rho K}+\frac{g_{\rho K}}{s-M^2+iM\Gamma(s)},
 \label{bwadd}
 \end{eqnarray}
  where $g_{K*\pi}$ and $g_{\rho K}$ are complex couplings, and $M$ and $\Gamma(s)$ are the mass and width of $K_1(1400)$ with
the $s$-wave width given by
\begin{equation}
\Gamma(s)=\Gamma_0\frac{q(s)}{q_\mathrm{on}}\Theta(\sqrt{s}-M_{K^*}-M_{\pi}).
\end{equation}
$q(s)$ and $q_\mathrm{on}$ are calculated by
 \begin{equation}
 q(s)=\frac{\lambda^{1/2}(s,M^2_\pi,M^2_{K^*})}{2\sqrt{s}}\quad\mbox{and}\quad
 q_\mathrm{on}=\frac{\lambda^{1/2}(M^2,M^2_\pi,M^2_{K^*})}{2M}.
 \end{equation}

In our model, Eq.~(\ref{bwadd}), we have the following adjustable
parameters: $a$, $b$, $g_{K^*\pi}$, $g_{\rho K}$, $M$ and
$\Gamma_0$. In principle, $f$ and $a(\mu)$ can also be taken as free
parameters. In order to limit the number of free parameters, we have
adopted the following procedure:
\begin{enumerate}
\item Starting from the values used in Ref.~\cite{Roca:2005nm},
we readjust $f$ slightly so that the higher pole position is close
to the experimental $K_1(1270)$. This gives $f$ a value of $\sim115$\,MeV.
\item Since we have a global arbitrary phase, we take $a$ real while $b$ is kept complex.
\item $M$ and $\Gamma_0$ are fixed at their experimental
values, i.e. $M=1402$\,MeV and $\Gamma=174$\,MeV. We note that a
reasonable readjustment of these two values only gives a slightly
better fit. Since this does not qualitatively improve our
interpretation of the data, we are satisfied with fixed $M$ and
$\Gamma$ (at the PDG values).

\item We minimize the difference between the WA3 data and our
calculated amplitudes to fix the other seven parameters.
\end{enumerate}

\begin{figure*}[ht]
\begin{center}
\begin{tabular}{cc}
\includegraphics[scale=0.4]{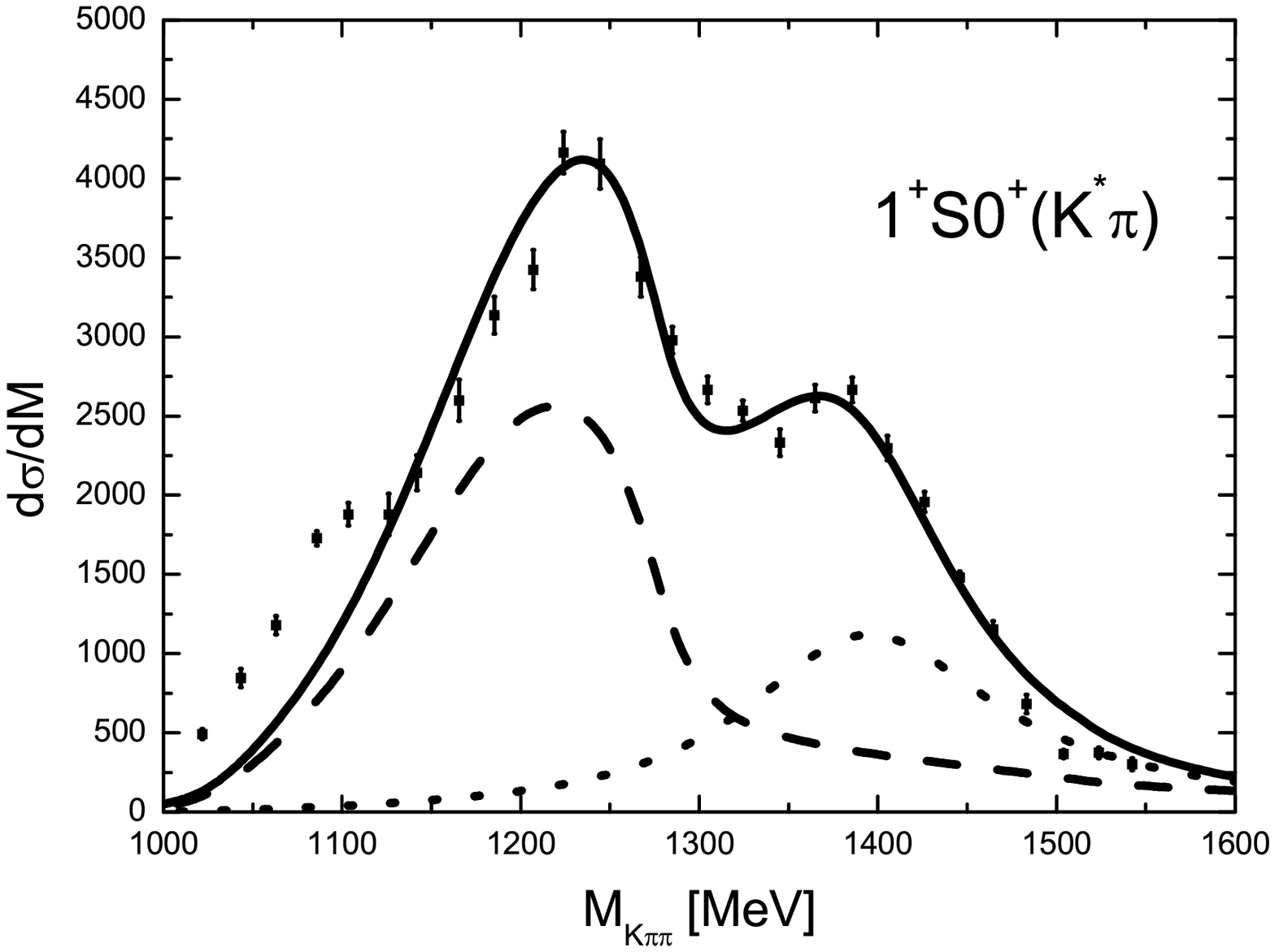} %
&\includegraphics[scale=0.4]{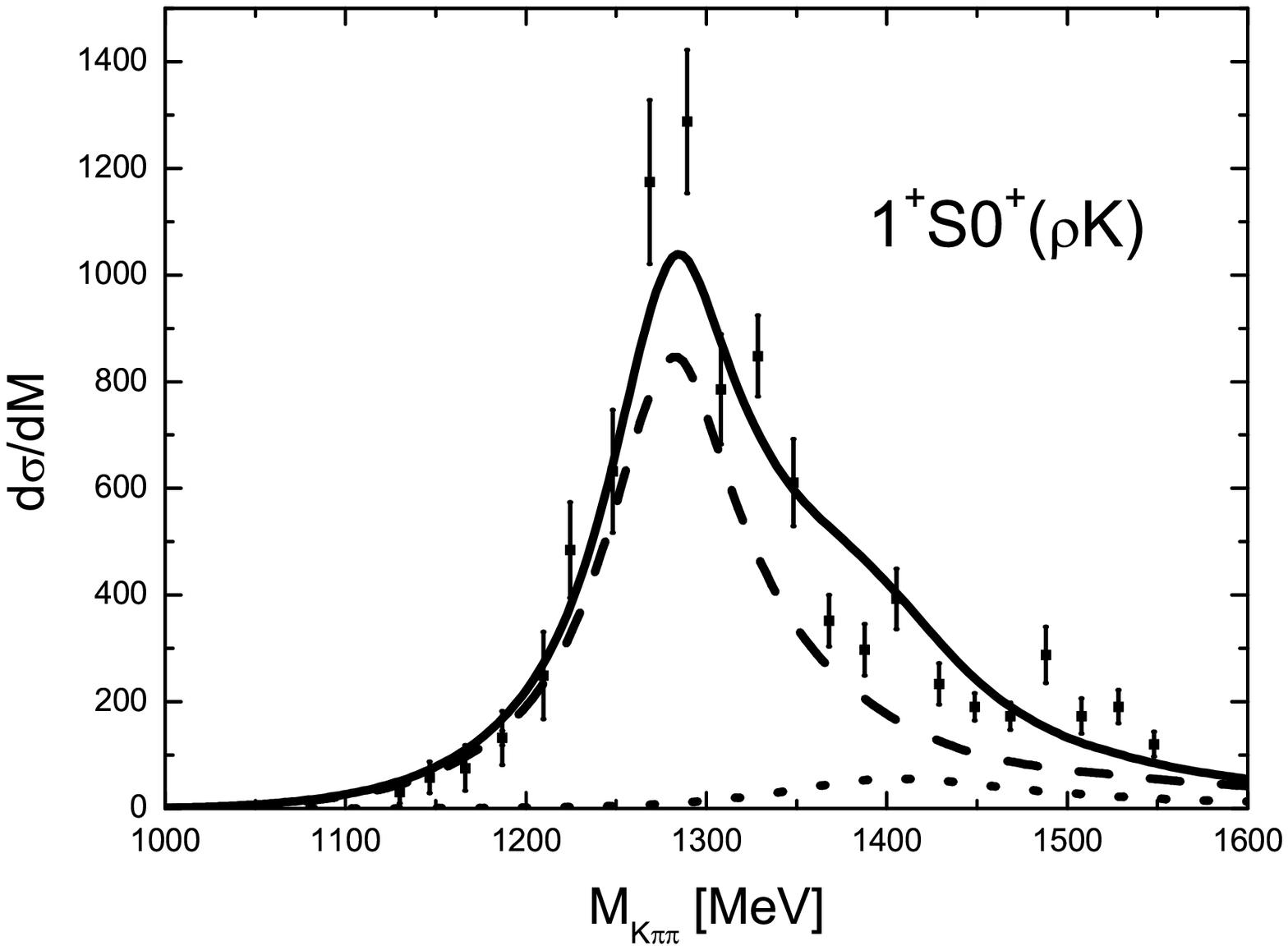}\\
\includegraphics[scale=0.4]{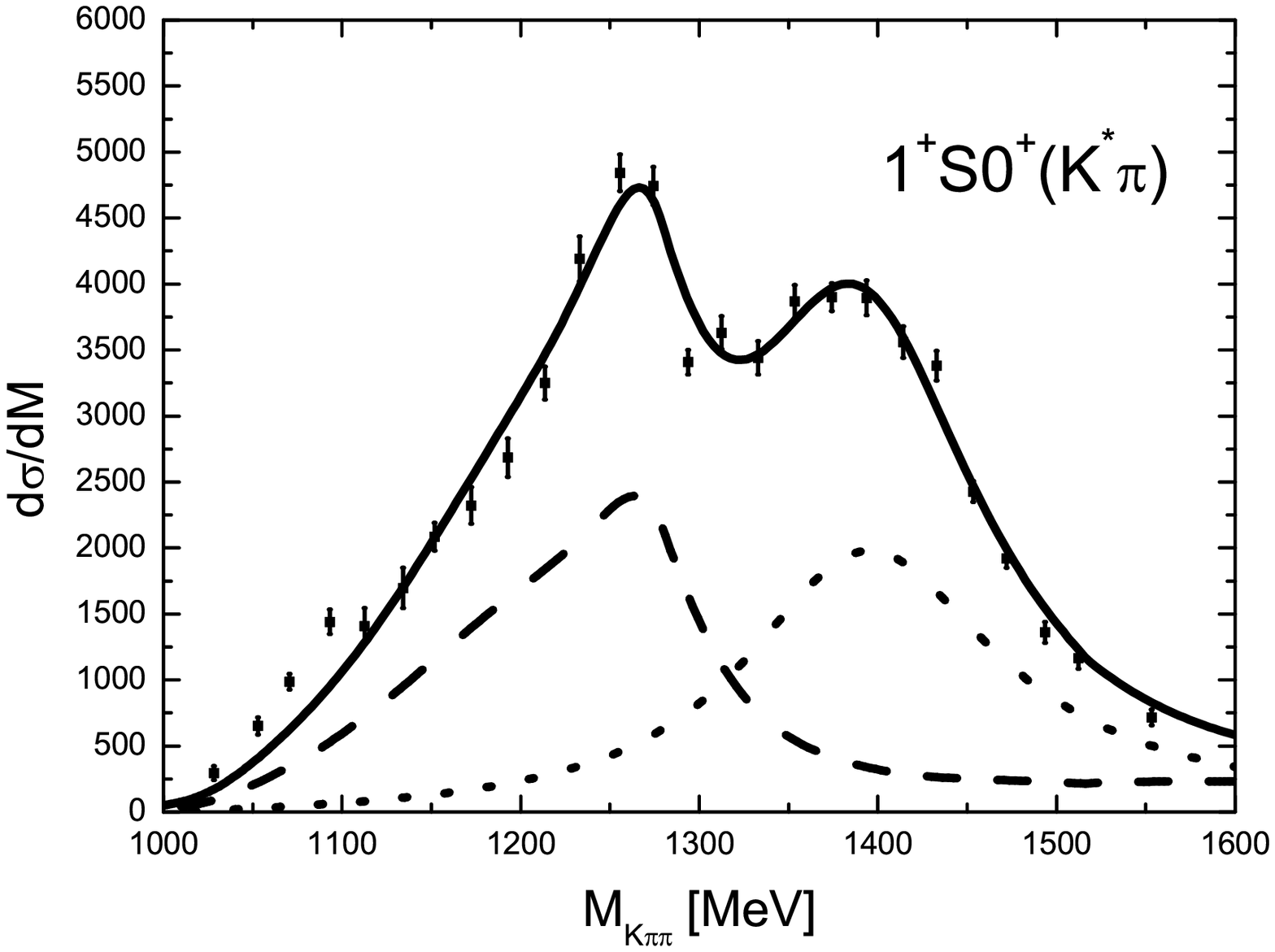}%
&\includegraphics[scale=0.4]{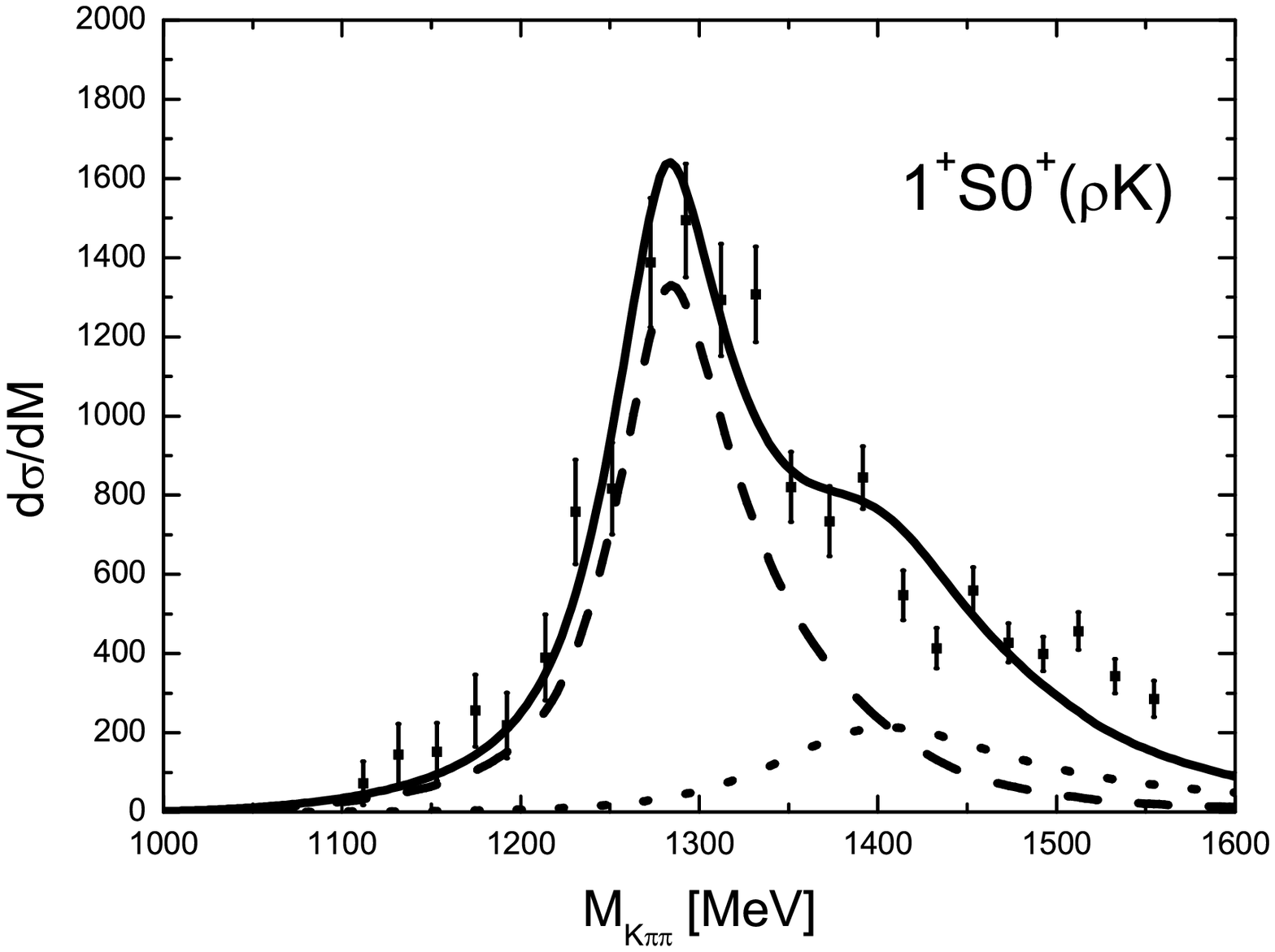}\\
\includegraphics[scale=0.4]{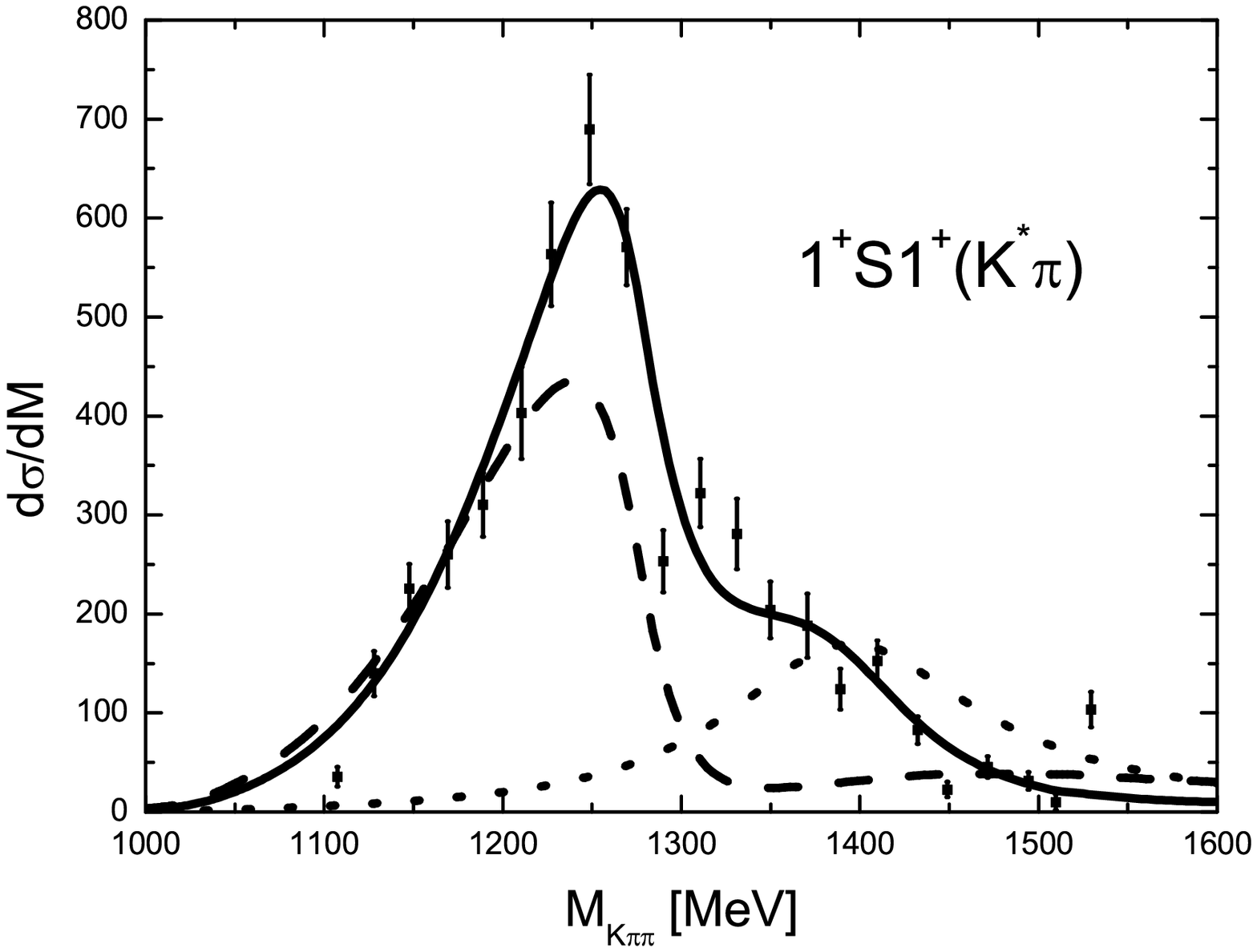}%
&\includegraphics[scale=0.4]{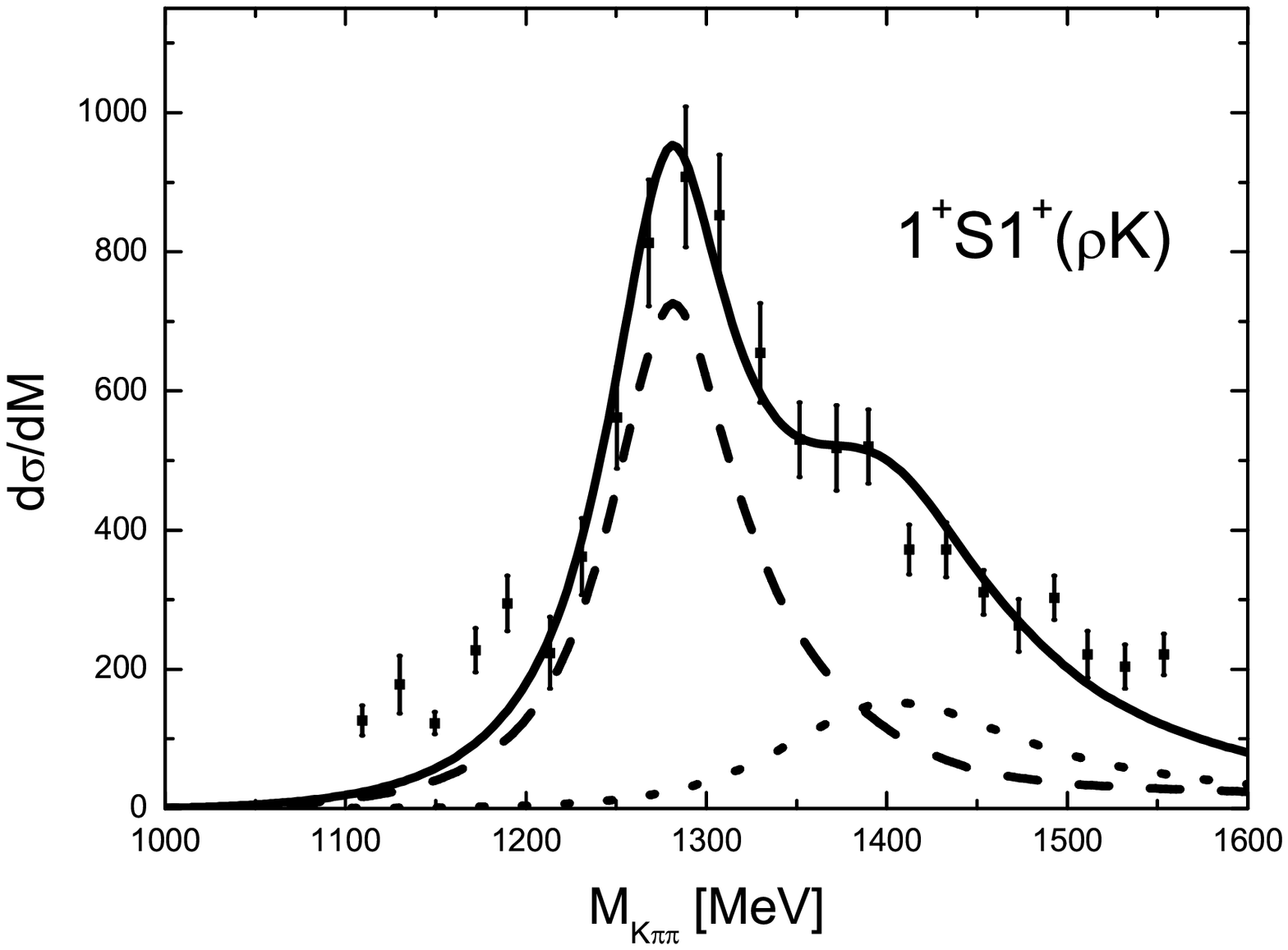}
\end{tabular}
\caption{$K^*\pi$ and $\rho K$ invariant mass distributions. The
data are from the WA3 reaction $K^- p\rightarrow K^-\pi^+\pi^- p$ at
63\,GeV~\cite{Daum:1981hb}. Data in the upper panels are for $0\le
|t'|\le 0.05$\,GeV$^2$ and those in the middle and bottom panels for
$0.05\le |t'|\le 0.7$\,GeV$^2$, where $t'$ is the four momentum
transfer squared to the recoiling proton. The data are further
grouped by $J^PLM^\eta$ followed by the isobar and odd particle. $J$
is the total angular momentum, $P$ the parity, $L$ the orbital
angular momentum of the odd particle. $M^\eta$ denotes the magnetic
substate of the $K\pi\pi$ system and the naturality of the
exchange.}
\label{fig:s1}
\end{center}
\end{figure*}
The results are shown in Fig.~\ref{fig:s1} in comparison with the
WA3 data~\cite{Daum:1981hb}. According to Ref.~\cite{Flatte:1976xu},
for a $s$-wave resonance, the theoretical differential cross section
can be calculated by
 \begin{equation}
 \frac{d\sigma}{dM}=c|T|^2q
 \end{equation}
 where $M$ is the invariant mass of the $K^*\pi$ or $\rho K$ systems,
 $c$ is a normalization constant, $T$ is the amplitude specified above for the $K^*\pi$ or $\rho K$
 channels
  and $q$ is the center of mass three-momentum  of $K^*\pi$ or $\rho K$. We have taken $c$ to be 1, or in other words, it has been
absorbed into the coupling constants $a$, $b$, $g_{K^*\pi}$ and
$g_{\rho K}$, which are tabulated in Table~\ref{tab:param}.
\begin{table*}[ht]
	\setlength{\tabcolsep}{0.4cm}
	\renewcommand{\arraystretch}{1.2}
\caption{Parameter values obtained from fitting the WA3
data~\cite{Daum:1981hb}. Data set 1, 2, 3 correspond to the low
$|t'|$ $1^+S0^+$ data, the high $|t'|$ $1^+S0^+$ data and the high
$|t'|$ $1^+S1^+$ data. The fits are performed by assigning an equal
error of 100 (events) to each data point. }
\begin{center}
\begin{tabular}{c|c|c|c|c}
\hline\hline &&&& \\[-4.5mm]
 Data set & $a$ & $b$ & $g_{K*\pi}$ (MeV$^2$) & $g_{\rho K}$ (MeV$^2$)  \\
\hline  &&&& \\[-4.5mm]
 1& 1.65 &($-1.60,0.27$)&($-221029,-341404)$&($-82631,-76119)$\\
 2& $-1.41$&($0.31,-2.06)$&($536803,64738)$&($-13388,-219728)$\\
 3& 0.45&($-1.32,-0.24)$&($109352,-114341)$&($-179938,-46491)$
 \\ \hline\hline
\end{tabular}
\label{tab:param}
\end{center}
\end{table*}
 From
Fig.~\ref{fig:s1},
 it is clearly seen that our model can fit the data around
the peaks very well. In Fig.~\ref{fig:s1},
 the dashed and dotted lines are the
separate contributions of $K_1(1270)$ and $K_1(1400)$. One can
easily see that $K_1(1400)$ decays dominantly to $K^*\pi$, which is
consistent with our present understanding of this
resonance~\cite{Yao:2006px}.

It should be mentioned that in our model the lower peak observed in
the invariant mass distribution of the $K^*\pi$ channel is due to
the contribution of the two poles of $K_1(1270)$. This is very
different from the traditional interpretation. For example, the
lower peak observed in the $K^*\pi$ invariant mass distributions of
$K^\pm p\rightarrow K^\pm \pi^+\pi^- p$ at 13\,GeV was interpreted
as a pure Gaussian background by Carnegie et
al.~\cite{Carnegie:1976cs}, which has a shape similar to the
contribution of the $K_1(1270)$ as shown in Fig.~\ref{fig:s1}.
 On the other
hand, the K-matrix approach was adopted to analyze the
SLAC~\cite{Brandenburg:1975gv} and the WA3 data~\cite{Daum:1981hb}.
In this latter approach, the lower peak mostly comes from the
so-called Deck background, which after unitarization, also has a
shape of resonance. As we mentioned in the introduction, even in the
original
 WA3 paper~\cite{Daum:1981hb}, it was noted that their model failed to
describe the $1^+S1^+$($K^*\pi$) data, in the notation $J^P L
M^\eta$ with $\eta$ the naturality of the
exchange~\cite{Daum:1981hb}. The predicted peak is 20\,MeV higher
than the data. If the fit were done only to the $K^*\pi$ data, the
agreement was much better but then the predicted $K_1(1270)$ would
be lower by 35\,MeV than that obtained when other channels were also
considered in the fit. We will discuss more about the K-matrix
approach in the following section.

It is worth stressing that the $K_1(1270)$ peak seen in the
upper-left panel of Fig.~\ref{fig:s1}
 is significantly broader than that in the
upper-right panel. Furthermore the peak positions are also different
in the two cases (1240\,MeV and 1280\,MeV respectively). Both
features have a straightforward interpretation in our theoretical
description since the first one is dominated by the low-energy
(broader) $K_1(1270)$ state, while the second one is dominated by
the higher-energy (narrower) $K_1(1270)$ state.

\begin{figure*}[ht]
\begin{center}
\includegraphics[scale=0.6]{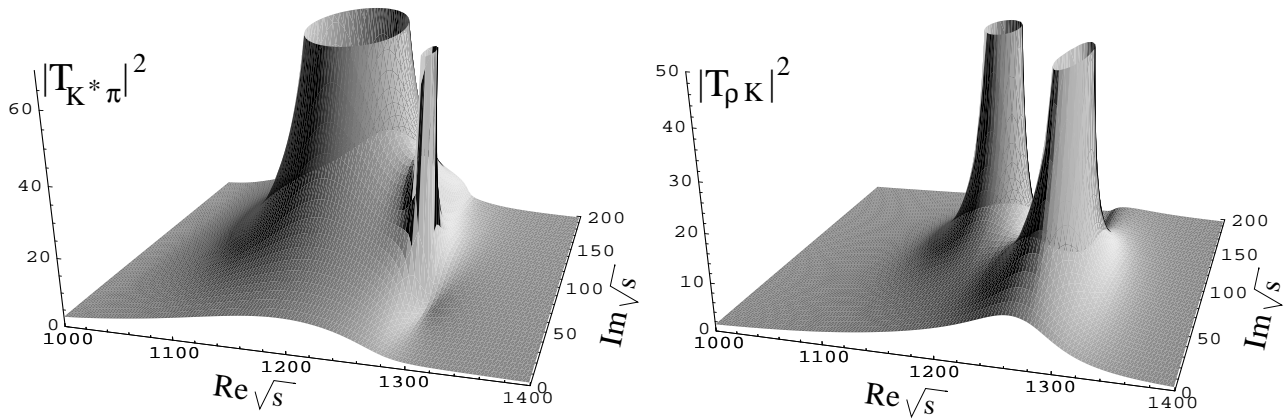} %
\caption{Modulus squared of the amplitudes of Eq.~(\ref{eq:Ts})
in the
unphysical Riemann sheet of the complex $\sqrt{s}$ variable. }
\label{fig:poles3D}
\end{center}
\end{figure*}
In order to see more clearly the contribution of the two
$K_1(1270)$ poles to the different reactions  ($K^*\pi$ and $\rho
K$), we show in Fig.~\ref{fig:poles3D} the modulus squared of the
amplitudes of Eq.~(\ref{eq:Ts}) in the unphysical Riemann sheet of
the complex $\sqrt{s}$ variable. The plots have been done with the
result  of the fit for the  $0\le |t'|\le 0.05$\,GeV$^2$ data. (The
other sets of data give analogous results). The relevant thing for
the evaluation of the  cross sections of the different reactions is
the value  of the amplitude in the real axis. We can see very
clearly the two different  $K_1(1270)$ poles and how their
different strength and position in the complex plane affects the
value in the real axis. For the $K^*\pi$ channel we clearly see
that the shape in the real axis is essentially determined by the
lower mass  pole, the higher one having a negligible  effect
despite being closer to the real axis. For the $\rho K$ case, the
shape in the real axis is mainly determined by the higher mass
pole. The lower mass pole has a minor influence. In the $\rho K$
case both poles have relevant strength but the fact that the lower
mass pole is far away from the real axis makes its effect on it
less relevant. It is also worth stressing that the shape of the
amplitude in the real axis differs from a Breit-Wigner--like
shape.

In a less microscopic approach than the one we do, 
the $K_1(1270)$ could also be parameterized  as two explicit Breit-Wigner contributions 
 in order to mimic the two poles building up the $K_1(1270)$, similarly as  done 
  in Eq.~(\ref{bwadd}) for the $K_1(1400)$ with only one Breit-Wigner.
However, this procedure requires the knowledge of the couplings  
to the main channels ($K^*\pi$ and $\rho K$), the masses and the widths of the two poles. 
In this way, one would require four complex parameters for the couplings and four real ones 
for the masses and widths, instead of just the two complex parameters actually 
used in Eq.~(\ref{eq:Ts}) for the $K_1(1270)$. Therefore, by removing one global phase,
 we would  be left 
with 13 free parameters, instead of just the three ones in Eq.~(\ref{eq:Ts}). 
 This large reduction in the number of free parameters is a remarkable advantage 
of  employing U$\chi$PT  in order to describe the two poles that
 build up the $K_1(1270)$. 
 Indeed, as discussed, 
the data are already well reproduced within our scheme, which explicitly generates  
the two poles associated with the $K_1(1270)$ \cite{Roca:2005nm},
 and adding 10 more free parameters would certainly obscure   
 any possible conclusion. Apart from that, the amplitudes in Eq.~(\ref{eq:Ts}) also 
contain non-resonant contributions, beyond what would be obtained by simply taking 
two Breit-Wigner poles so as to give the amplitudes. In addition, the reproduction of the 
WA3 data poses an intriguing test to the amplitudes of Ref.~\cite{Roca:2005nm},
 which is one of our main aims  as well.

\section{K-matrix approach}

Since the PDG data are largely based on the WA3
data~\cite{Daum:1981hb}, it seems worthwhile to look more closely at
the model employed by C. Daum et al. to derive $K_1(1270)$ and
$K_1(1400)$ properties~\cite{Daum:1981hb}. A detailed description of
the model can be found in Refs.~\cite{Daum:1981hb,Daum:1980ay,
Bowler:1976qe,Bowler:1974th,Bowler:1975my}. Here, we only briefly
summarize the relevant formulae. The production amplitudes are given
by
\begin{equation}\label{eq:pa}
F=(1-iK\rho)^{-1}P,
\end{equation}
where $K$ is a $n\times n$ matrix with $n$ the number of channels
taken into account. In Eq.~(\ref{eq:pa}), $\rho$ is a diagonal matrix
consisting of phase space and its elements are
 \begin{equation}
 \rho(s)_{ij}=\frac{p_i}{8\pi\sqrt{s}}\delta_{ij}.
  \end{equation}
 The $K$ matrix element is of
the following form:
\begin{equation}
K_{ij}=\frac{f_{ai}f_{aj}}{M_a-M}+\frac{f_{bi}f_{bj}}{M_b-M},
\end{equation}
where the decay couplings $f_{ai}$ and $f_{bi}$ are assumed to be
real numbers. The production vector $P$ consists of the Deck
amplitudes $D$ and the direct production terms $R$
\begin{equation}
P=(1+\alpha K)D+R.
\end{equation}
The Deck amplitudes $D$ are parameterized by
 \begin{equation}
  D_i=D_{i0}e^{i\phi_i}/(M^2_{K\pi\pi}-M^2_K)
  \end{equation}
  and $R$ by
  \begin{equation}
  R_i=\frac{f_{pa}f_{ai}}{M_a-M}+\frac{f_{pb}f_{bi}}{M_b-M},
  \end{equation}
  where the production couplings $f_{pa}$ and $f_{pb}$ are complex
  numbers~\cite{Daum:1981hb,Daum:1980ay}. Furthermore, one can assume $f_{pa}$ to be
real. For the
  constant $\alpha$, a value of $0.4$ was used in
  Ref.~\cite{Daum:1981hb}. The authors commented that the final results
  do not depend sensitively on this value. As we will see below,
  this might not be the case.

  Five channels have been included in the ACCMOR
  analysis~\cite{Daum:1981hb}. For the sake of simplicity, as in other similar theoretical
  analyses~\cite{Carnegie:1976cs,Bowler:1976qe}, we have
  used only two channels, i.e. $K^*\pi$ and $\rho K$, which are the most relevant in the analysis of Ref.~\cite{Daum:1981hb}.
   In this case, there are 13
  free parameters: 4 for decay couplings, 3 for production couplings, 4 for
  Deck backgrounds, and 2 for K-matrix poles $M_a$ and $M_b$. When
  this model is used to study the high $|t'|$ WA3 data, the following
  assumption is used, i.e. the decay couplings are the same for $1^+S0^+$
  and $1^+S1^+$ data, but the production couplings and Deck backgrounds can
  be different. Therefore, for high $|t'|$ data, there are 20 free
  parameters.

\begin{figure*}[ht]
\begin{center}
\begin{tabular}{cc}
\includegraphics[scale=0.4]{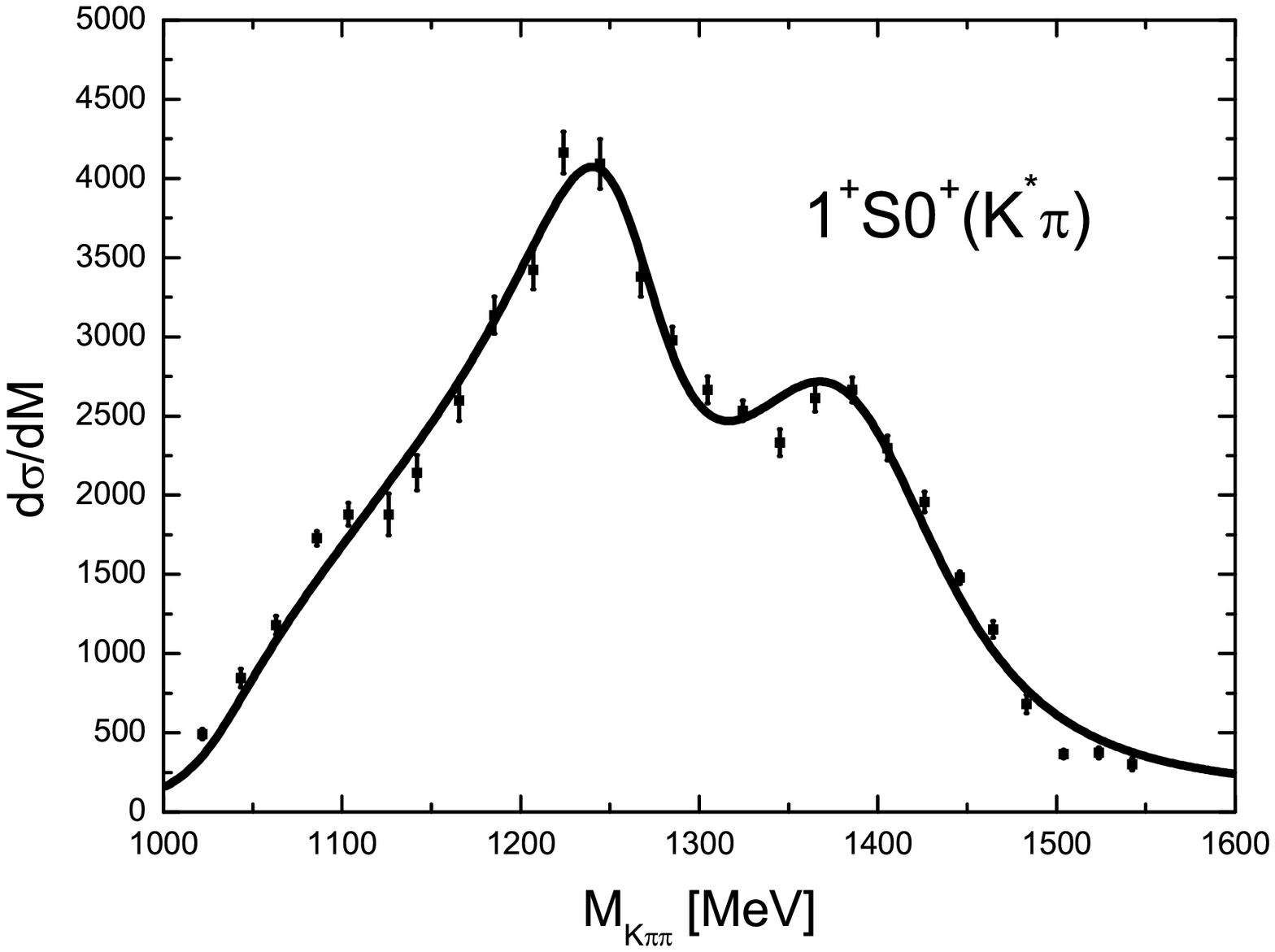} %
&\includegraphics[scale=0.4]{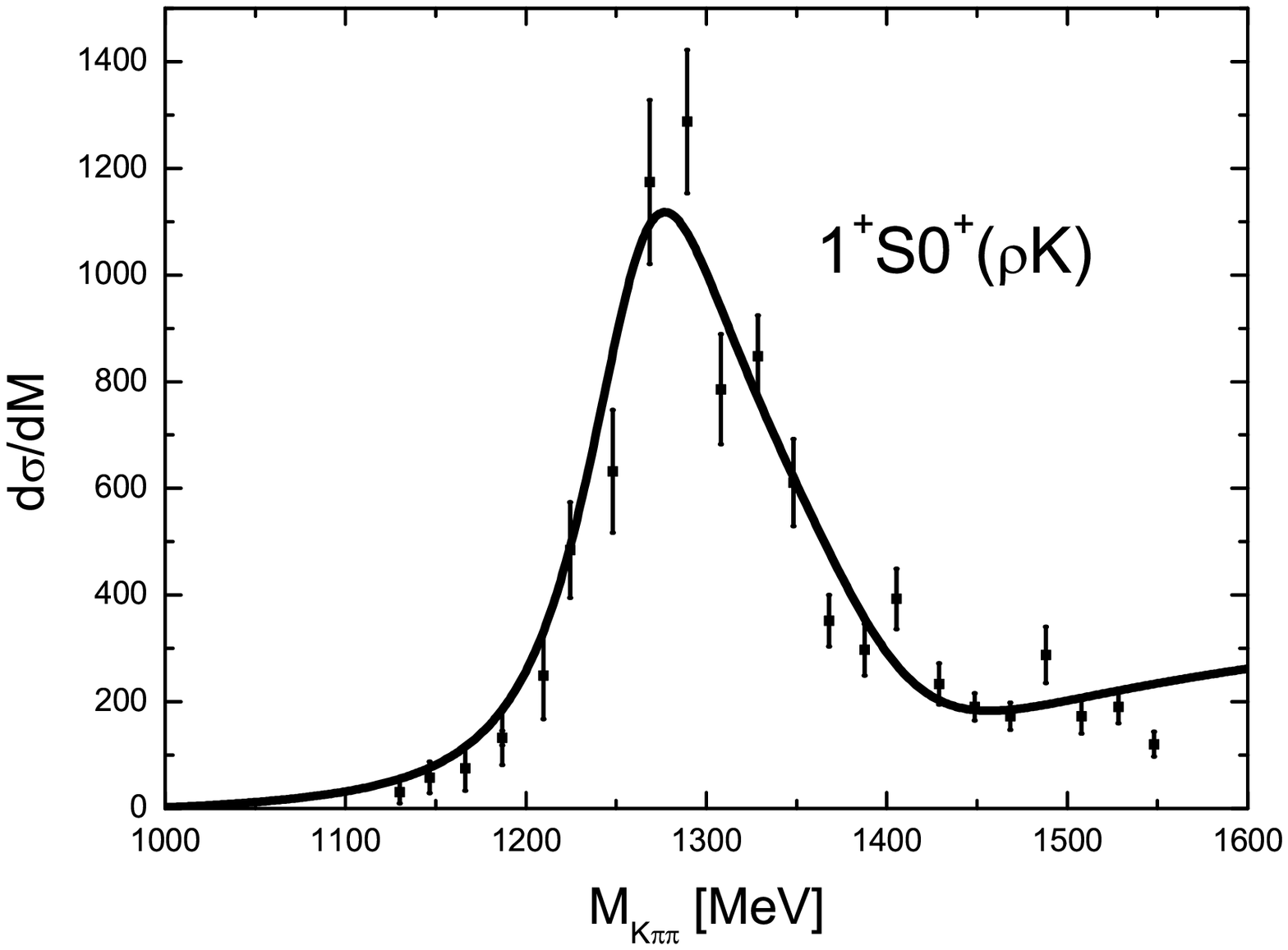}\\
\includegraphics[scale=0.4]{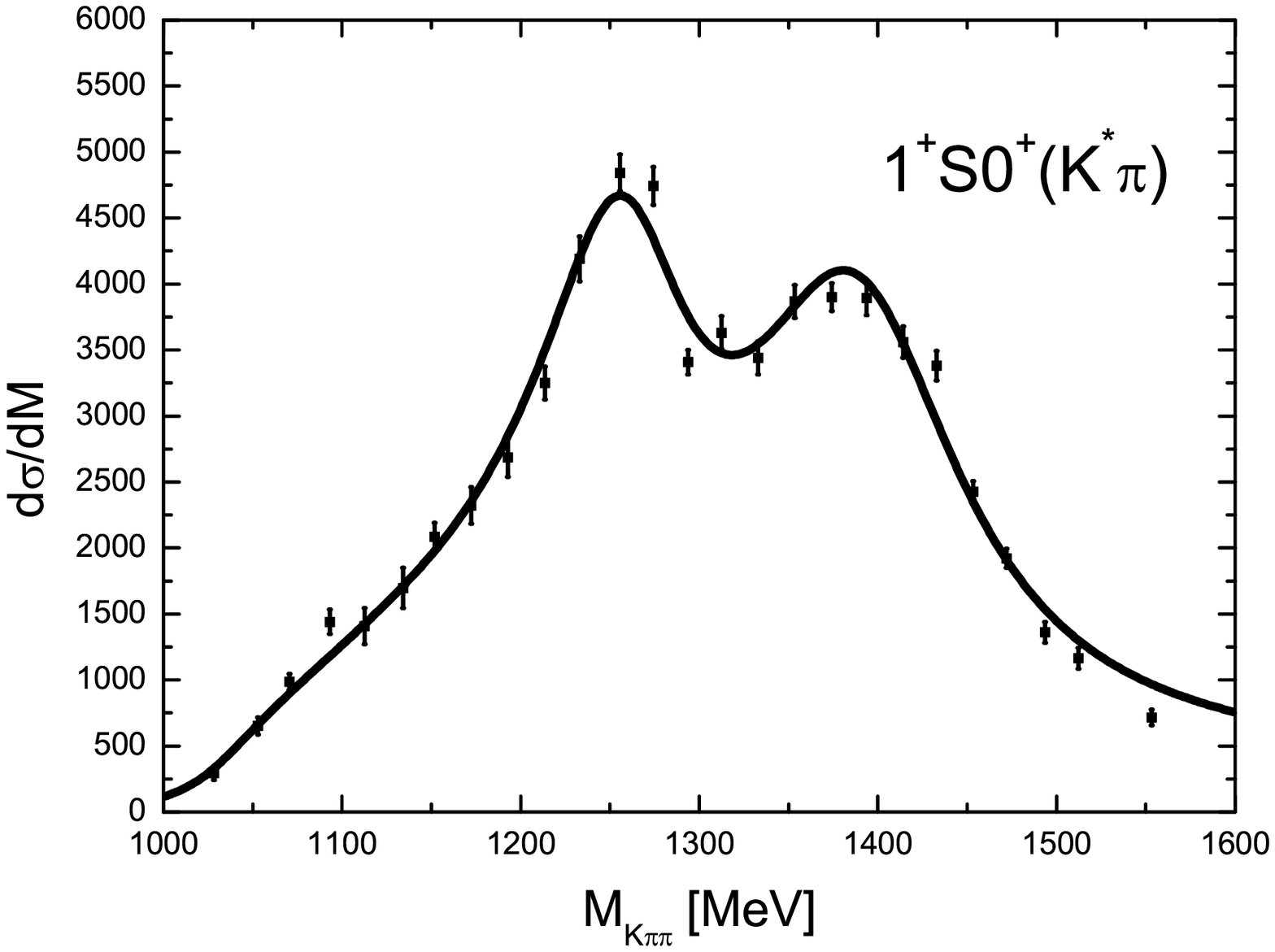}%
&\includegraphics[scale=0.4]{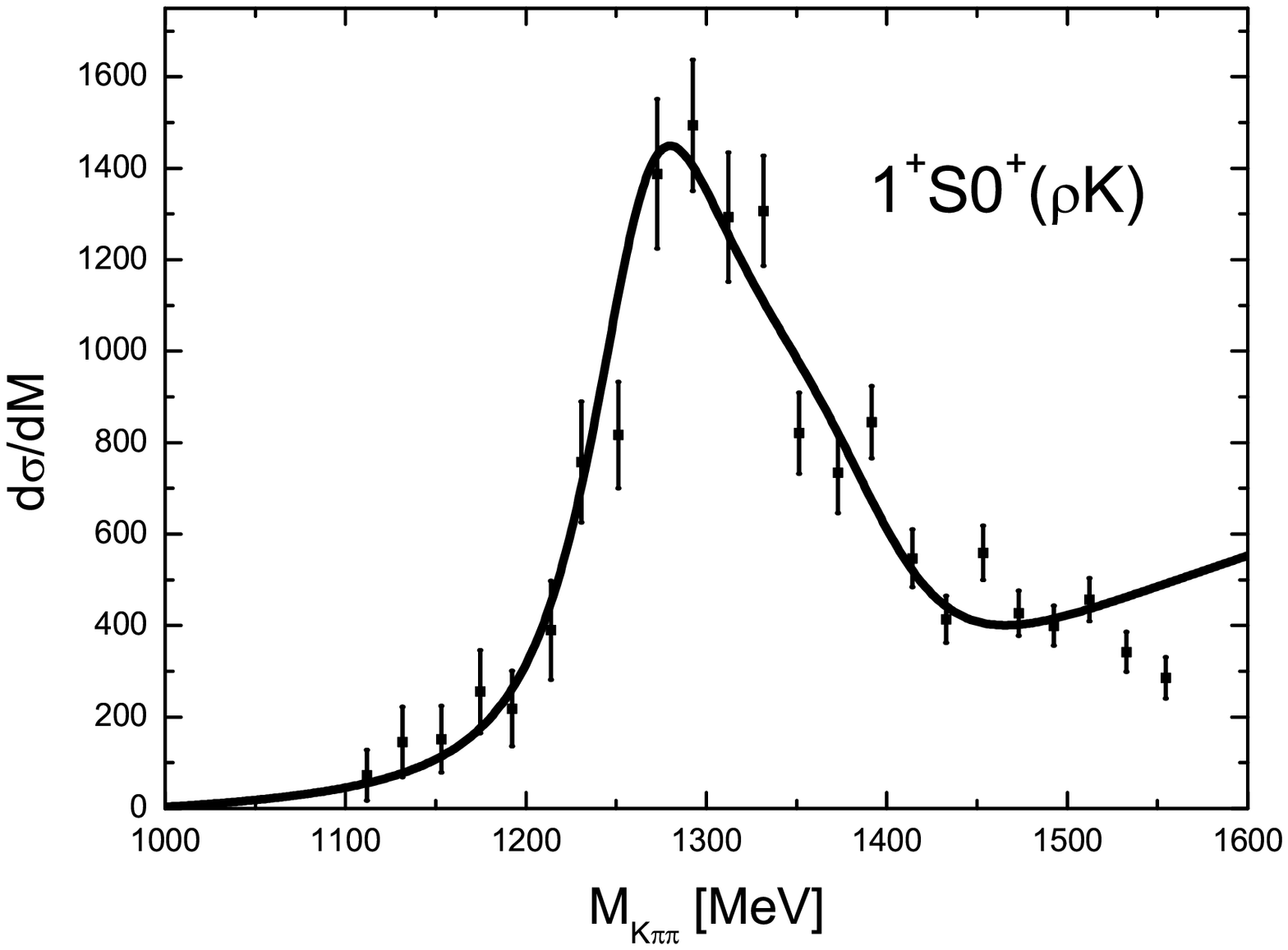}\\
\includegraphics[scale=0.4]{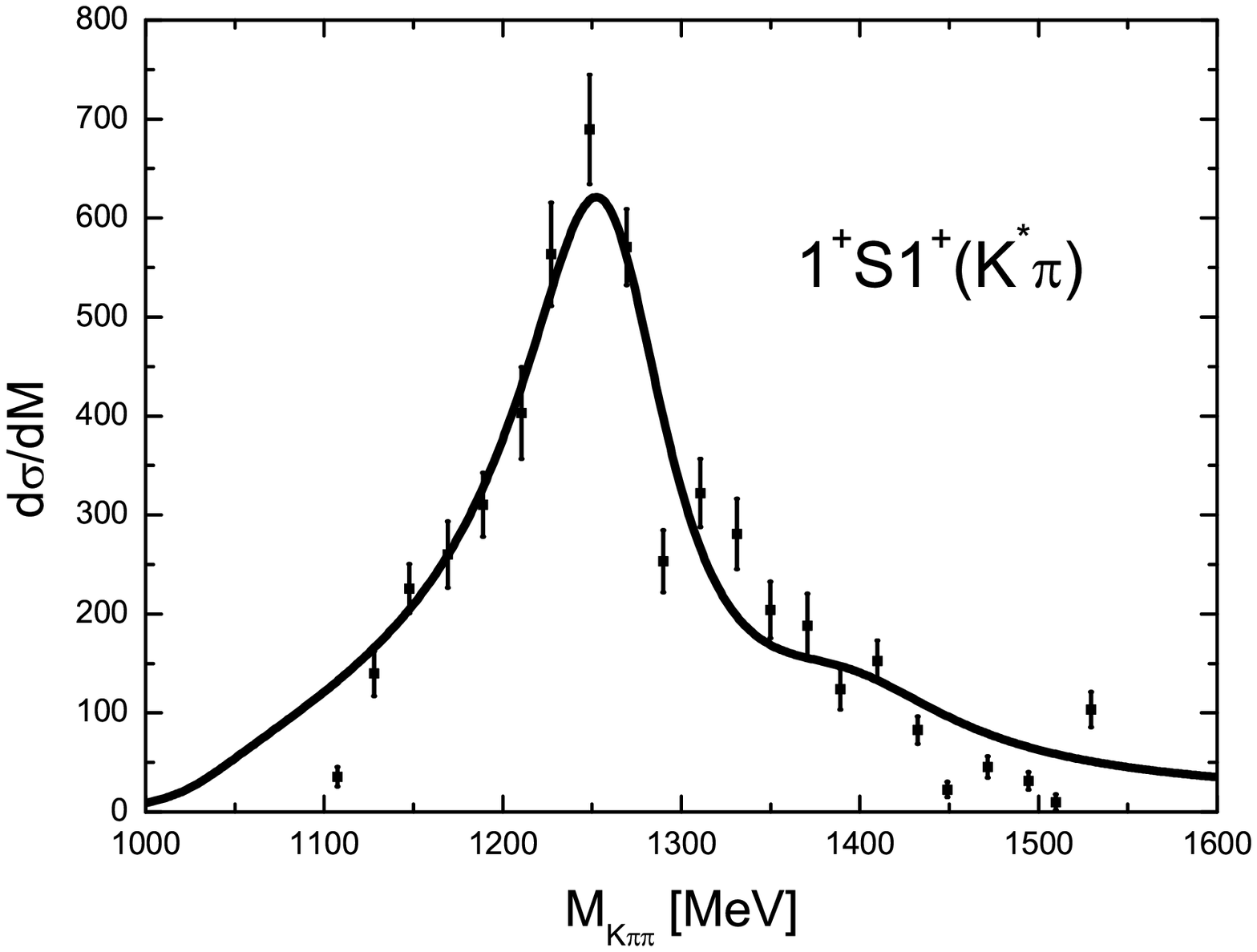}%
&\includegraphics[scale=0.4]{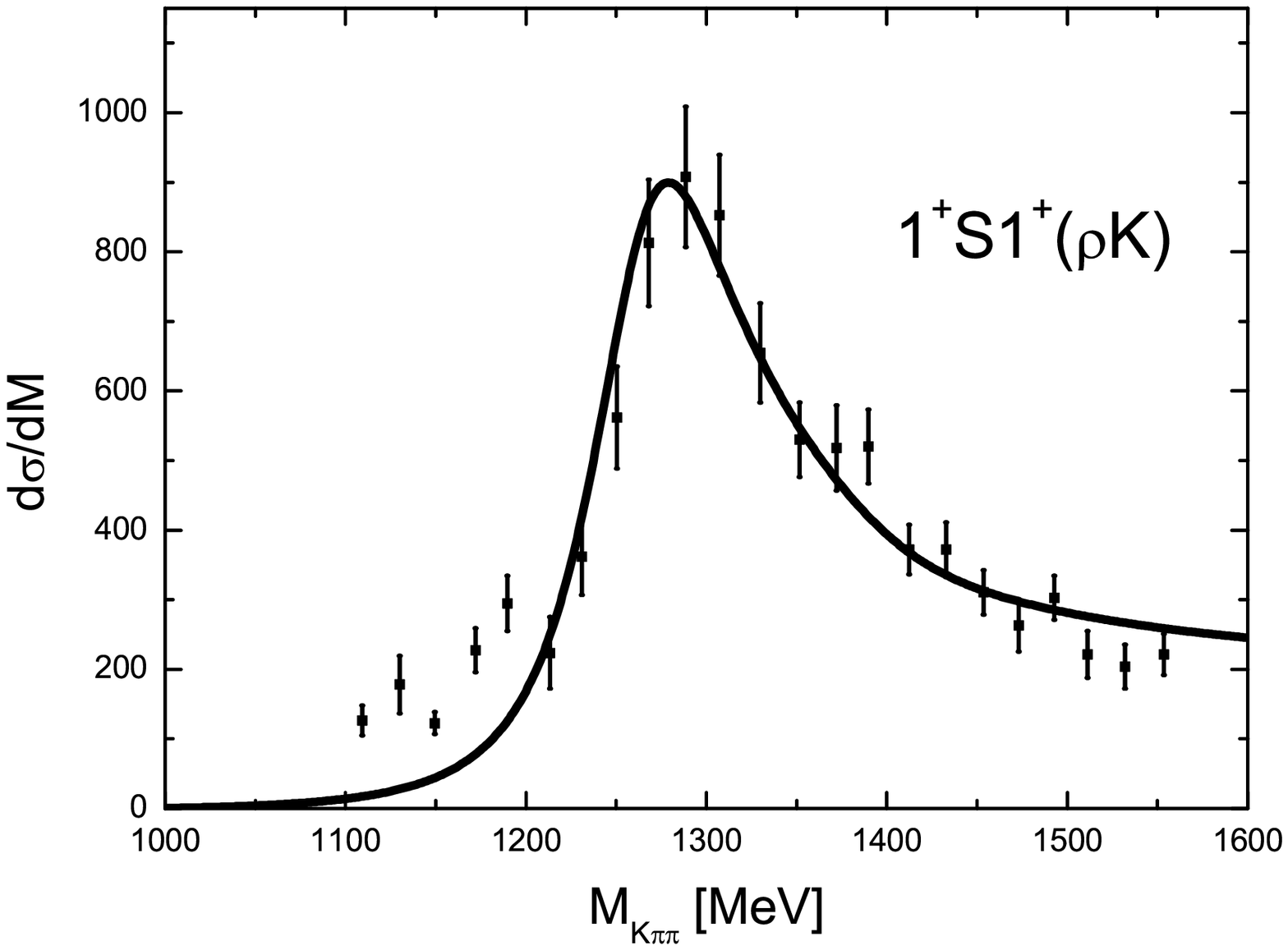}\\
\end{tabular}
\caption{The WA3 data in contrast with the K-matrix approach fit
with $M_a=1400$\,MeV, $M_b=1170$\,MeV and $\alpha=0.4$.}
\label{fig:CERN}
\end{center}
\end{figure*}
  In our fit of the WA3 data, we fix $M_a$ at 1400\,MeV  and $M_b$ at 1170\,MeV following
  the ACCMOR analysis~\cite{Daum:1981hb}. Thus, for low $|t'|$ data, we have 11 free
  parameters and for high $|t'|$ data, we have 18 parameters. The
  obtained results are contrasted with the WA3 data in
  Fig.~\ref{fig:CERN}.
   It is
  seen that the agreement is remarkably good. However, one should
  keep in mind the following: (i) Compared to our model, one has
  more freedoms in the K-matrix approach. (ii) Although the fit using
  the K-matrix approach is quantitatively better than our method, the
  fits are qualitatively very similar.

\begin{figure*}[ht]
\begin{center}
\begin{tabular}{cc}
\includegraphics[scale=0.4]{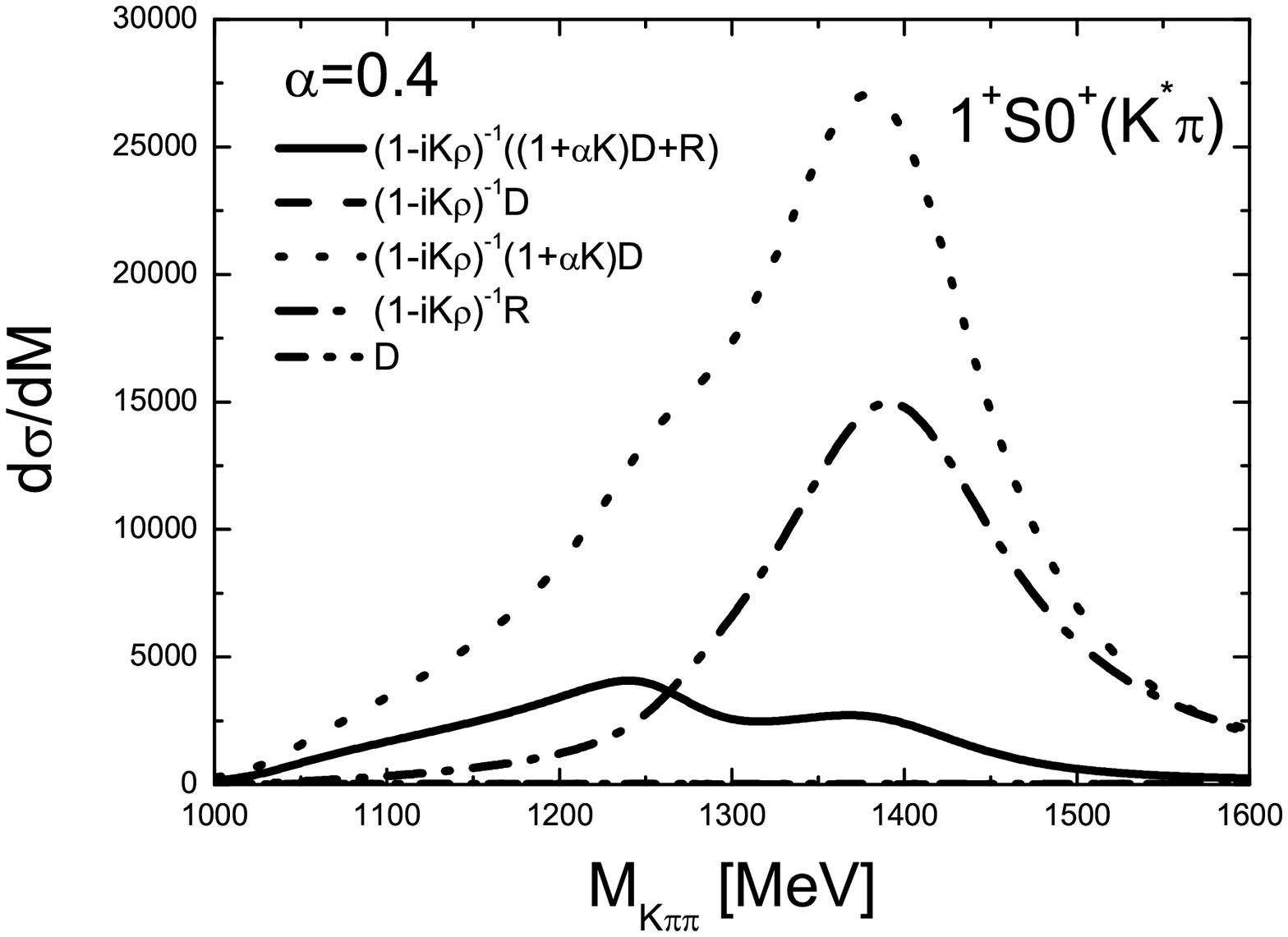} %
&\includegraphics[scale=0.4]{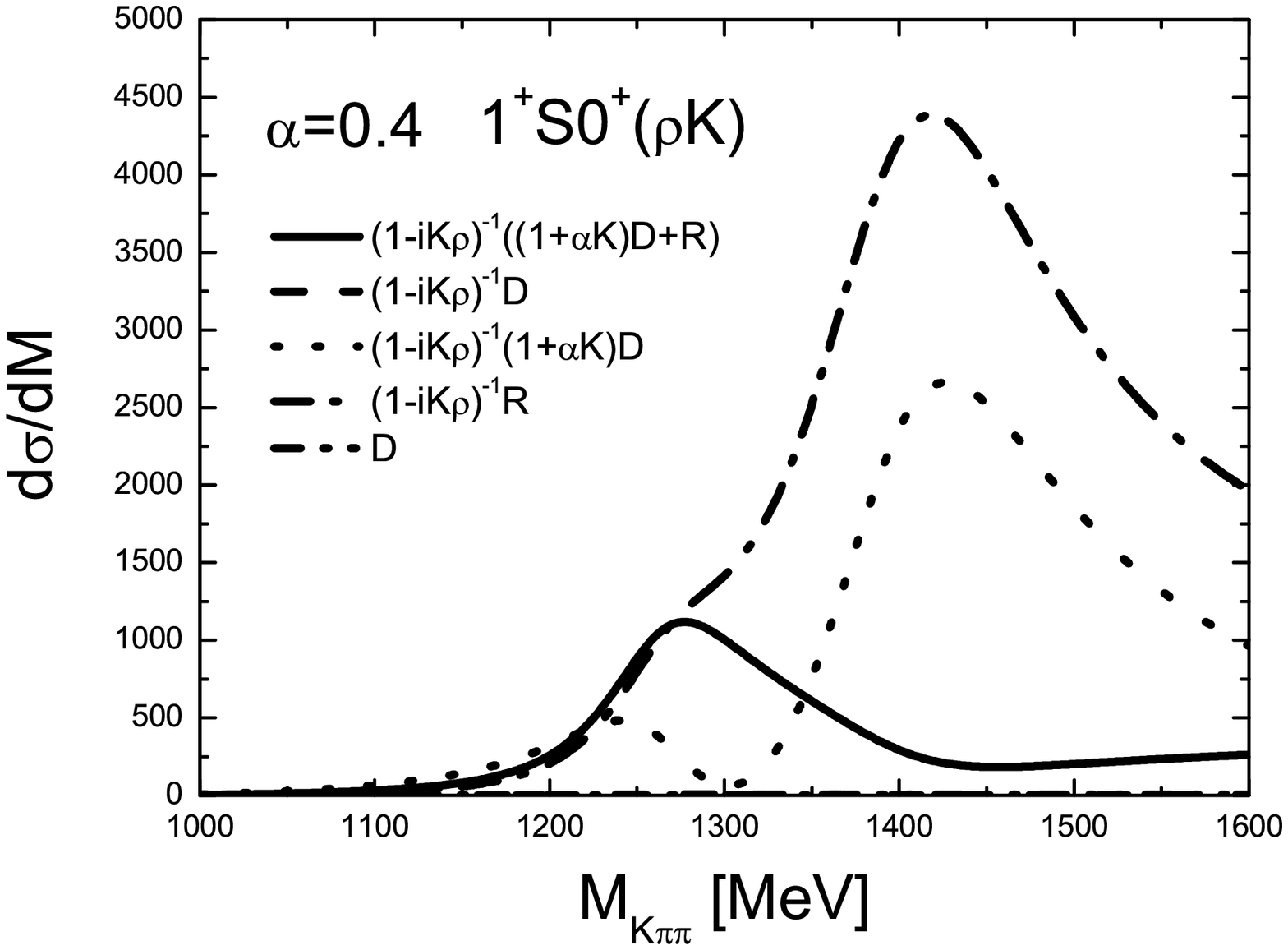}\\
\includegraphics[scale=0.4]{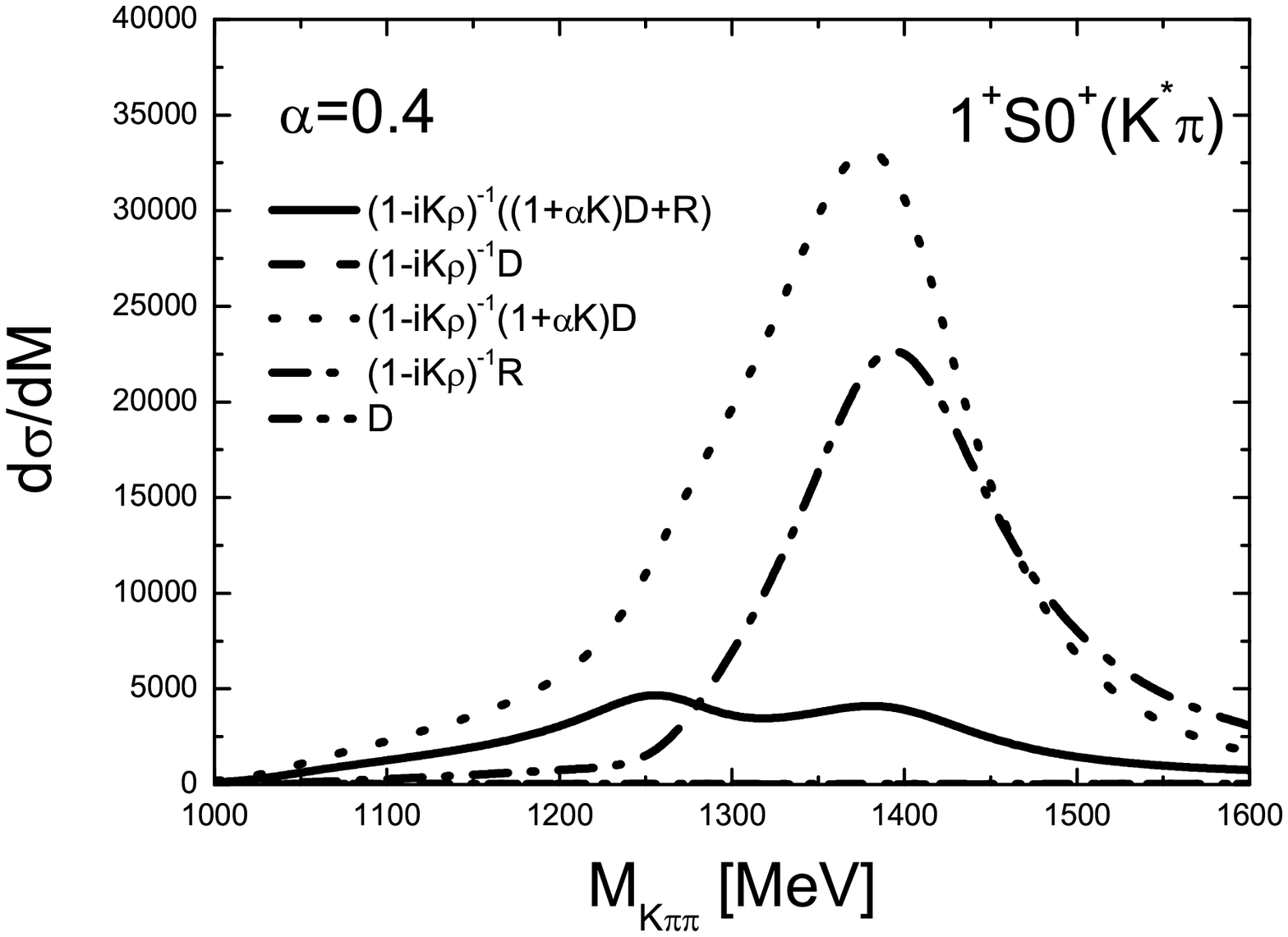}%
&\includegraphics[scale=0.4]{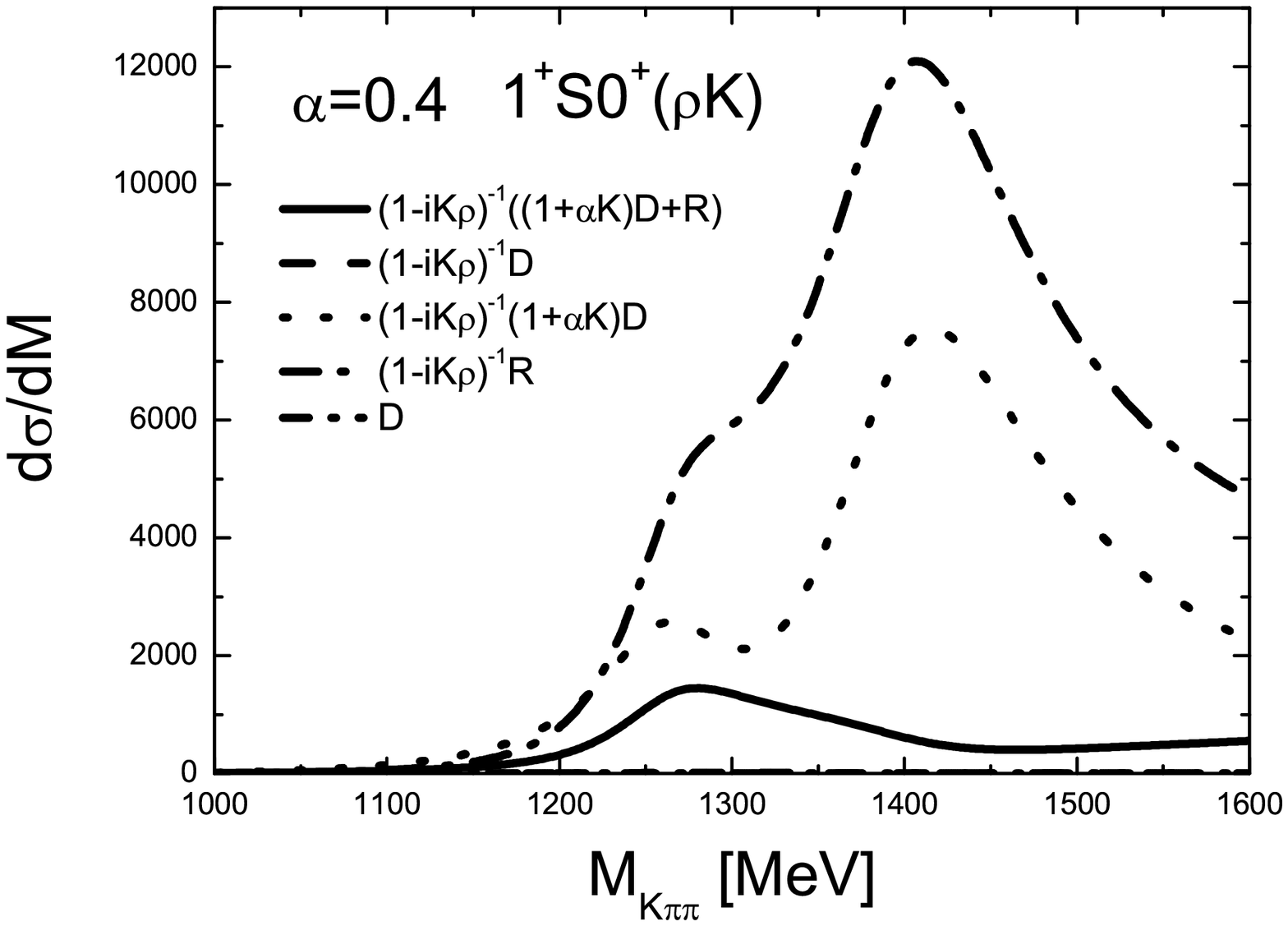}\\
\includegraphics[scale=0.4]{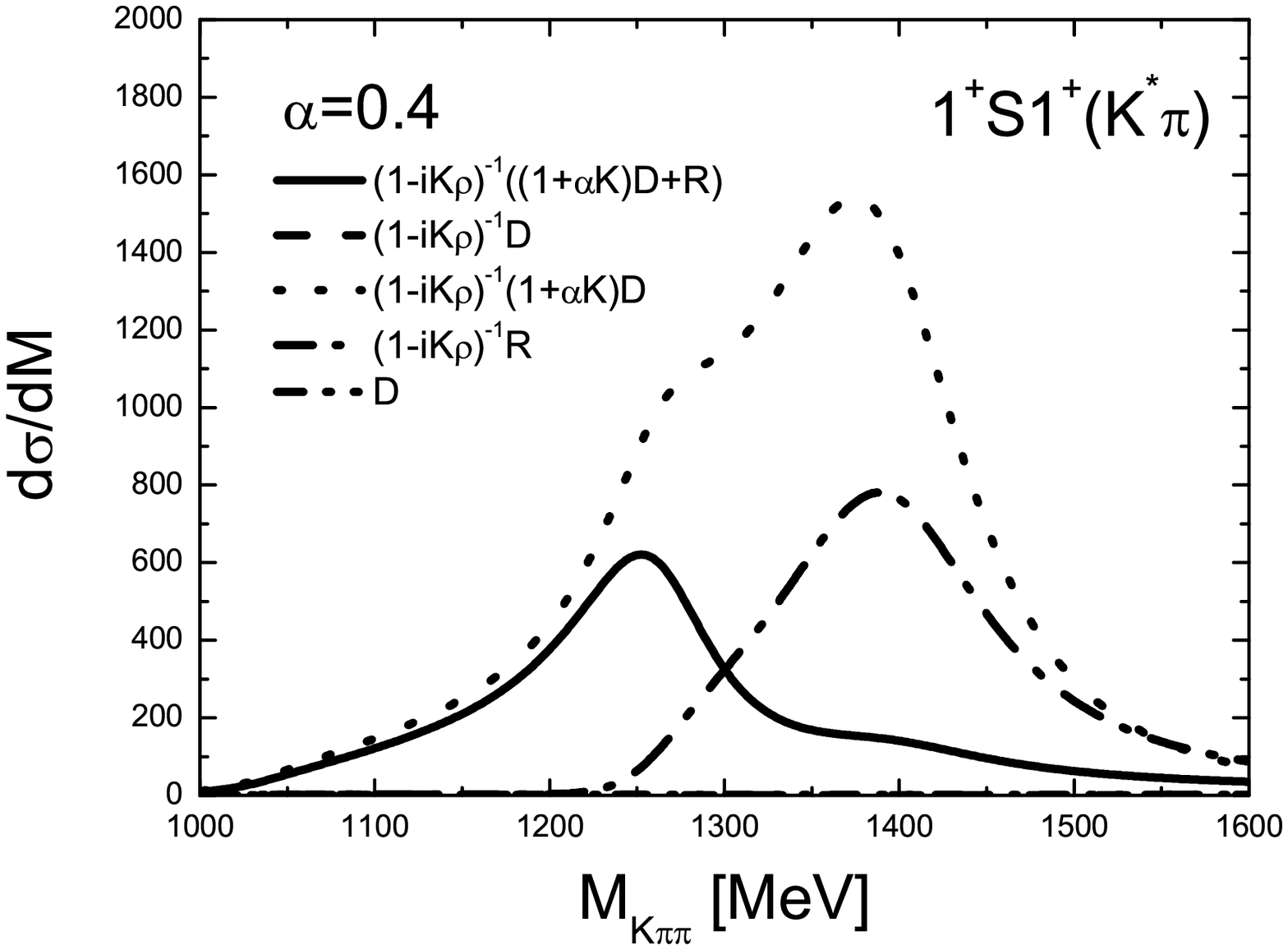}%
&\includegraphics[scale=0.4]{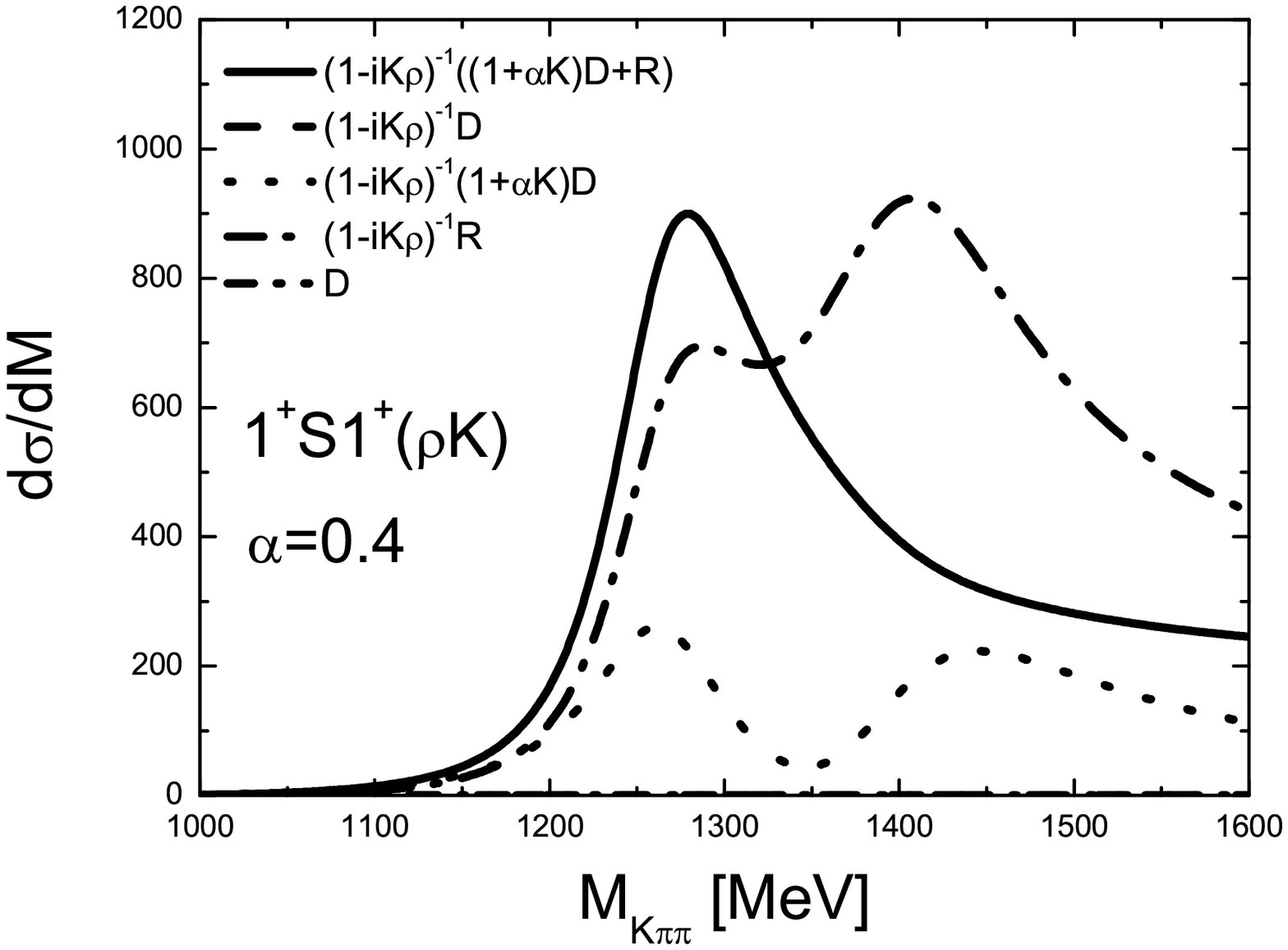}\\
\end{tabular}
\caption{The same as Fig.~\ref{fig:CERN},
 but with the contribution of the
components of the production amplitude $F$ plotted.}
\label{fig:CERN2}
\end{center}
\end{figure*}
  Now, we would like to study the contributions of the
  components of the production amplitudes $F$ of Eq.~(\ref{eq:pa}). In particular, we
  want to understand the contribution of Deck backgrounds.
   In Fig.~\ref{fig:CERN2},
    we plot the following quantities $|T_i|^2 q$ with $T_i$ being
  one of the following:
  \begin{eqnarray}
  T_1&=&(1-iK\rho)^{-1}((1+\alpha K)D +R)\quad\mbox{the full
  amplitude}\nonumber\\
  T_2&=&(1-iK\rho)^{-1}D\quad\mbox{the unitarized Deck
  background}\nonumber\\
  T_3&=&(1-iK\rho)^{-1}(1+\alpha K)D\quad\mbox{the full
  background}\nonumber\\
  T_4&=&(1-iK\rho)^{-1}R\quad\mbox{the direct production
  amplitude}\nonumber\\
  T_5&=&D\quad\mbox{the pure Deck background}\nonumber
  \end{eqnarray}
  From Fig.~\ref{fig:CERN2},
  it is seen that only $T_3$ and $T_4$ have relevant
  contributions to the total amplitude. These are the full background
  and the direct production terms. Surprisingly, the total amplitude
  seems to originate from the cancellation of two extremely large components:
  the background and the direct production. In the $K^*\pi$ low
  $|t'|$ data, both the background and the direct production
  amplitudes show no sign of a peak at the experimental lower peak position. Therefore, the lower peak shown in the total amplitude
   is completely due to the delicate interference between these two
  large components. Such large destructive interferences seem to be rather artificial, 
  particularly when compared to our fits, where each peak corresponds to a resonance structure. 
 In this sense, our interpretation of the $K_1(1270)$ as two poles seems to be favoured, {\it 
 a priori}, over the just discussed K-matrix approach, although more data should be compared 
 to reach a definitive, physically sound, statement on this.

  In the above fit, we have fixed $\alpha$ at 0.4, the value used by
  the ACCMOR collaboration~\cite{Daum:1981hb}. In Ref.~\cite{Daum:1981hb}, it was
  mentioned that the fitted results do not depend sensitively on the
  value of $\alpha$. We, therefore, have refitted the data by
  taking $\alpha$ as 0.2 and 0.0. It was found that with smaller
  $\alpha$, the results obtained are still far away from a
  straightforward interpretation but less
  extreme than the case with $\alpha=0.4$. But even if we take
  $\alpha$ to be zero, the whole amplitude seems to be a result of a delicate interference between
  the background and the direct production amplitude. Thus, this seems to
  indicate that the predictions of the K-matrix approach are not very
  stable. Furthermore, the results with this approach do not stand a clean physical interpretation.
  We have also refitted the data by setting $M_b$ at 1230\,MeV and 1270\,MeV. The results are in general similar to
  those with $M_b=1170$\,MeV. The only difference is that for the $M_b=1270$\,MeV case, the
peak of the high $|t'|$ $K^*\pi$ results appear 20\,MeV higher than
the data, thus $M_b=1270$\,MeV seems to be excluded.

\section{Contribution of other channels}
In our empirical model we constructed above to analyze the WA3 data,
we have only included two channels, i.e. $K^*\pi$ and $\rho K$. They
are the most important channels as can be clearly seen from
Figs.~\ref{fig:tVPVP} and \ref{fig:GtVPVP}.
 On the other hand, since in our chiral unitary
approach, we have three extra channels: $\phi K$, $\omega K$ and
$K^*\eta$, it would be interesting to show explicitly that their
inclusion does not significantly modify our analysis and conclusion.
Of course, to include more channels implies, in principle, more free
parameters. To overcome this drawback, we turn to SU(3) relations.
Supposing that $K_1(1270)$ has octet quantum numbers of flavor, its
couplings to different channels can be obtained through the
following interaction Lagrangian:
 \begin{equation}
 \mathcal{L}=D\langle\{\phi,V\}S\rangle+F\langle[\phi,V]T\rangle,
\label{lags} \end{equation}
where $\phi$ is the pseudoscalar nonet, $V$ the vector octet, and
$T$ and $S$ are the octets with positive and negative charge
conjugation, respectively. Then the couplings to different channels
can be obtained as:
 \begin{eqnarray} g_{K^*\pi}=\langle
R|\tilde{t}|K^*\pi\rangle&=&\sqrt{\frac{3}{2}}(D+F)\nonumber\\
g_{\rho K}=\langle R|\tilde{t}|\rho K\rangle&=&-\sqrt{\frac{3}{2}}(D-F)\nonumber\\
g_{K^*\eta}=\langle
R|\tilde{t}|K^*\eta\rangle&=&\frac{1}{\sqrt{6}}(D-3F)\nonumber\\
g_{\omega K}=\langle
R|\tilde{t}|\omega K\rangle&=&-\frac{1}{\sqrt{2}}(D-F)\nonumber\\
g_{\phi K}=\langle R|\tilde{t}|\phi K\rangle&=&-(D+F)
\label{eqs}
 \end{eqnarray}

 We remark that the Lagrangian Eq.~(\ref{lags}) follows from just flavour SU(3) symmetry, 
 without invoking any chiral symmetry. Indeed, the relations in Eq.~(\ref{eqs}) can also be 
 obtained by simply applying the Wigner-Eckert theorem for SU(3)~\cite{Georgi}. 
 Of course, one should expect flavour SU(3) violations in the couplings,  but since they are 
 expected to be moderate, and the addition of the extra channels just gives rise
  to small corrections, these violations would have a very small effect in our fits. 
 
Our total production amplitudes for $K^*\pi$ and $\rho K$ are then
\begin{widetext}
\begin{eqnarray}
T_{K^*\pi}\equiv
T_{\bar{K}^{*0}\pi^-}&=&\sqrt{\frac{2}{3}}g_{K^*\pi}+\sqrt{\frac{2}{3}}g_{\phi
K}G_{\phi K}t_{\phi K\rightarrow K^*\pi}+\sqrt{\frac{2}{3}}g_{\omega
K}G_{\omega K}t_{\omega K\rightarrow K^*\pi}
+\sqrt{\frac{2}{3}}g_{\rho K}G_{\rho K}t_{\rho K\rightarrow K^*\pi}\nonumber\\
&&+\sqrt{\frac{2}{3}}g_{K^* \eta}G_{K^* \eta}t_{K^*\eta\rightarrow
K^*\pi}+\sqrt{\frac{2}{3}}g_{K^*\pi}G_{K^*\pi}t_{K^*\pi\rightarrow
K^*\pi}+\frac{\sqrt{\frac{2}{3}}g'_{K^*\pi}}{s-M^2+iM\Gamma(s)},
\end{eqnarray}
\begin{eqnarray}
T_{\rho K}\equiv T_{\rho^0 K^-}&=&-\sqrt{\frac{1}{3}}g_{\rho
K}-\sqrt{\frac{1}{3}}g_{\phi K}G_{\phi K}t_{\phi K\rightarrow \rho
K}-\sqrt{\frac{1}{3}}g_{\omega K}G_{\omega K}t_{\omega K\rightarrow
\rho K}
-\sqrt{\frac{1}{3}}g_{\rho K}G_{\rho K}t_{\rho K\rightarrow \rho K}\nonumber\\
&&-\sqrt{\frac{1}{3}}g_{K^* \eta}G_{K^* \eta}t_{K^*\eta\rightarrow
\rho K}-\sqrt{\frac{1}{3}}g_{K^*\pi}G_{K^*\pi}t_{K^*\pi\rightarrow
\rho K}-\frac{\sqrt{\frac{1}{3}}g'_{\rho K}}{s-M^2+iM\Gamma(s)},
\end{eqnarray}
\end{widetext}
where $g'_{K^*\pi}=\sqrt{\frac{3}{2}}(D'+F')$ and $g'_{\rho
K}=\sqrt{\frac{3}{2}}(F'-D')$ are the couplings of $K_1(1400)$ to
the $K^*\pi$ and $\rho K$ channels by using the same SU(3) symmetry
arguments for the $K_1(1400)$ resonance. The Clebsch-Gordan
coefficient $\sqrt{\frac{2}{3}}$($-\sqrt{\frac{1}{3}}$) again
accounts for isospin projection. It is also worth stressing that 
if the previous equation is particularized to the case of only two channels, 
$K^*\pi$ and $ \rho K$, it would be equivalent to Eq.~(\ref{bwadd}).

\begin{figure*}[ht]
\begin{center}
\begin{tabular}{cc}
\includegraphics[scale=0.4]{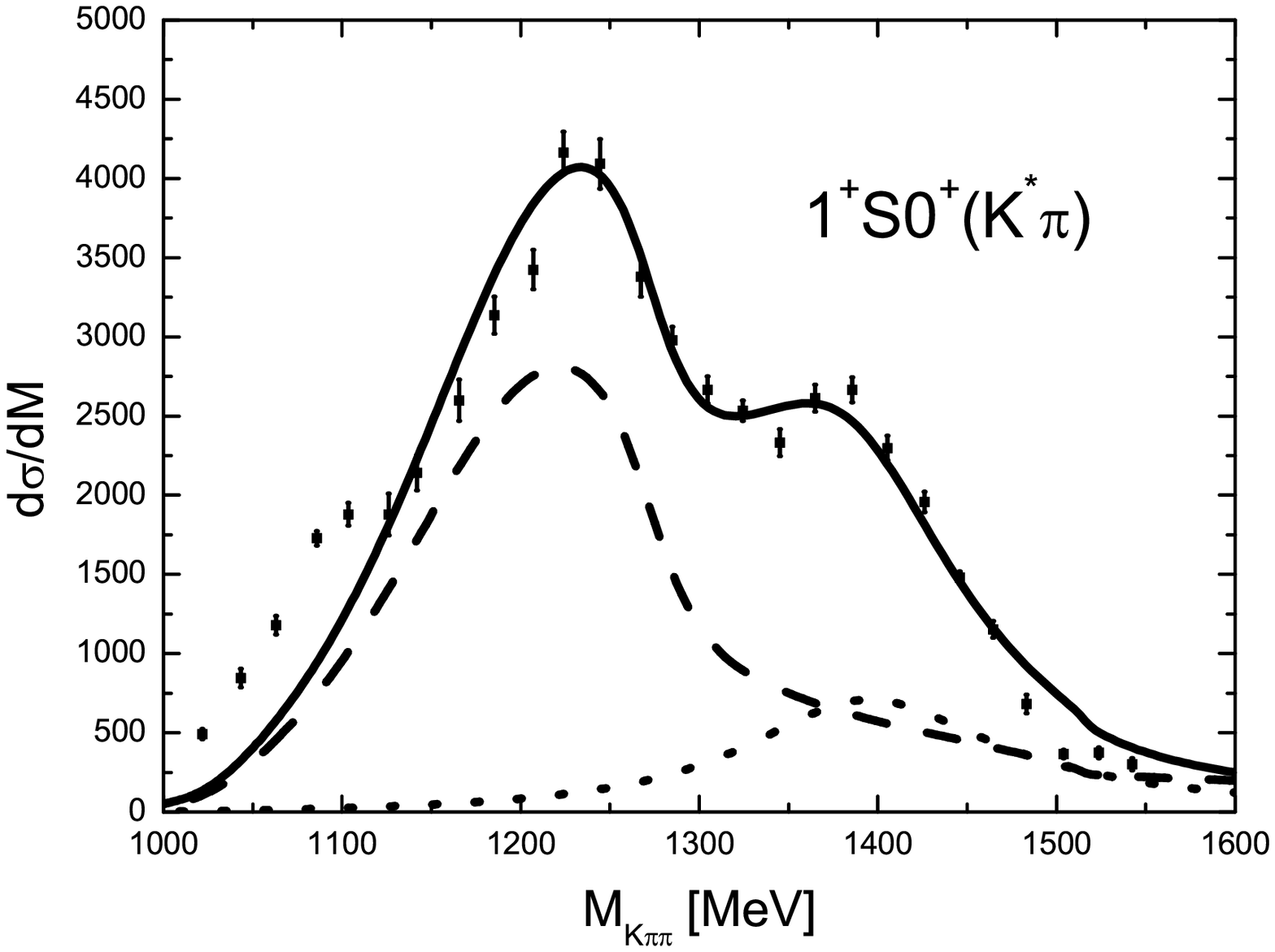} %
&\includegraphics[scale=0.4]{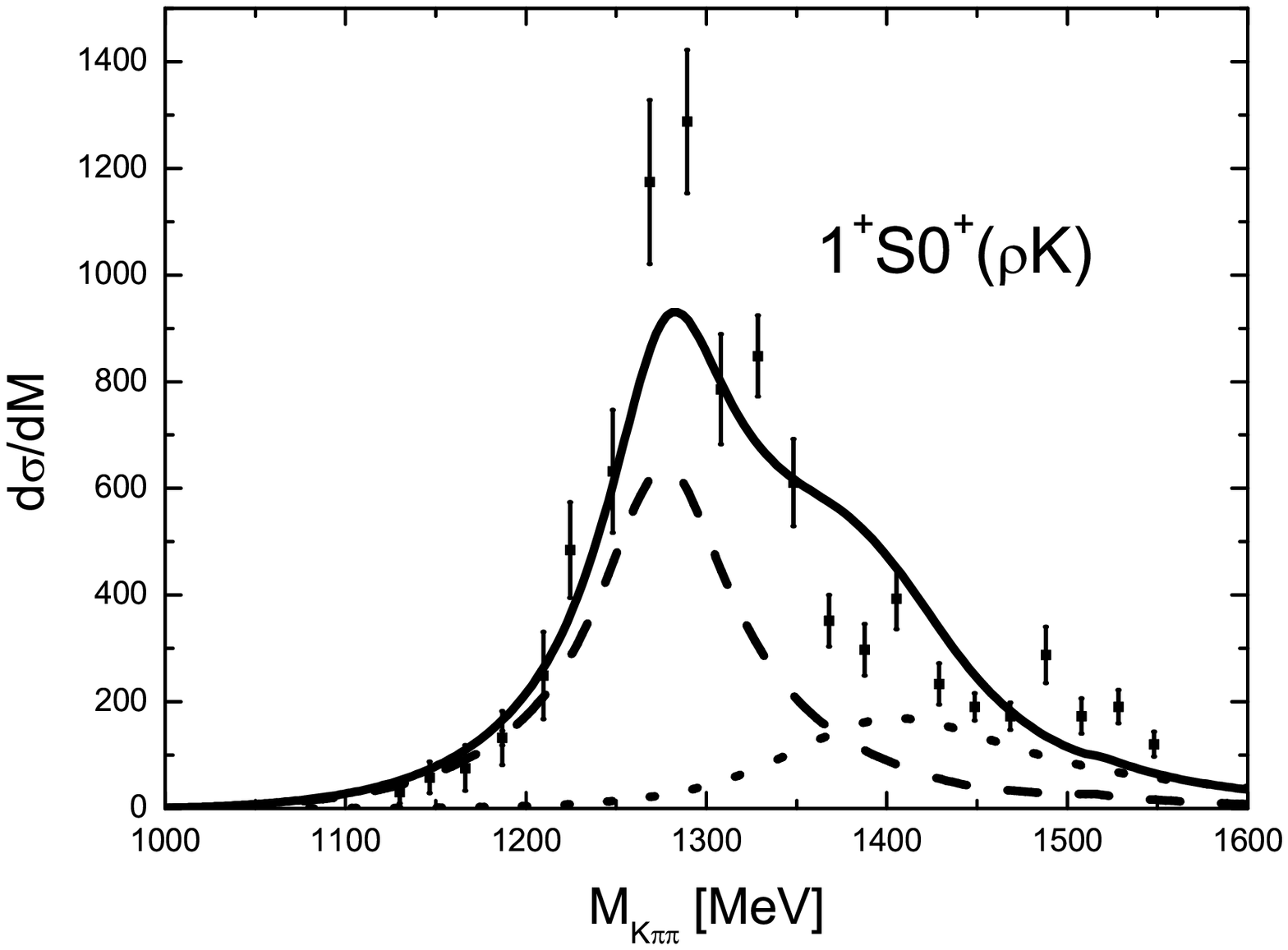}\\
\includegraphics[scale=0.4]{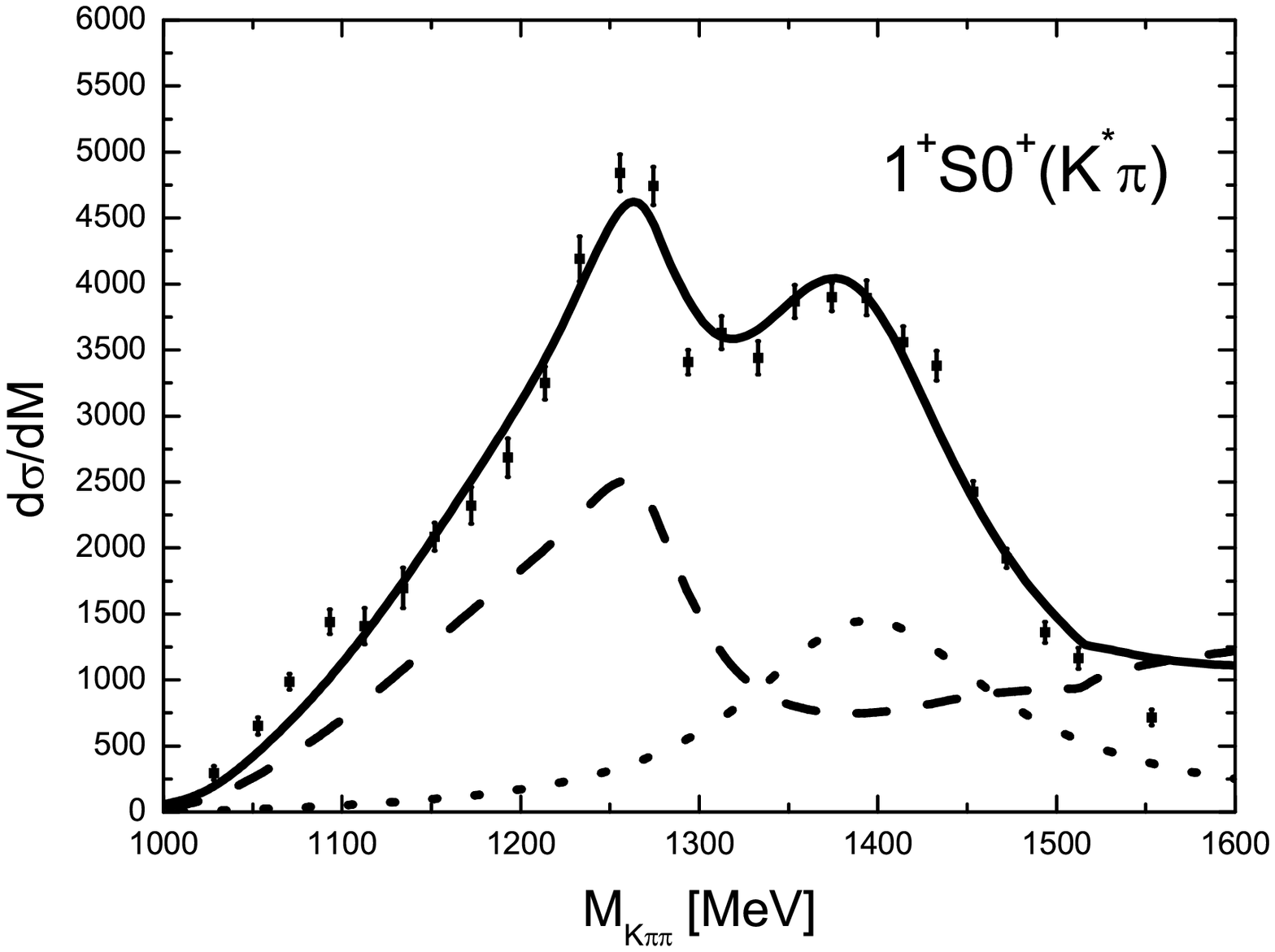}%
&\includegraphics[scale=0.4]{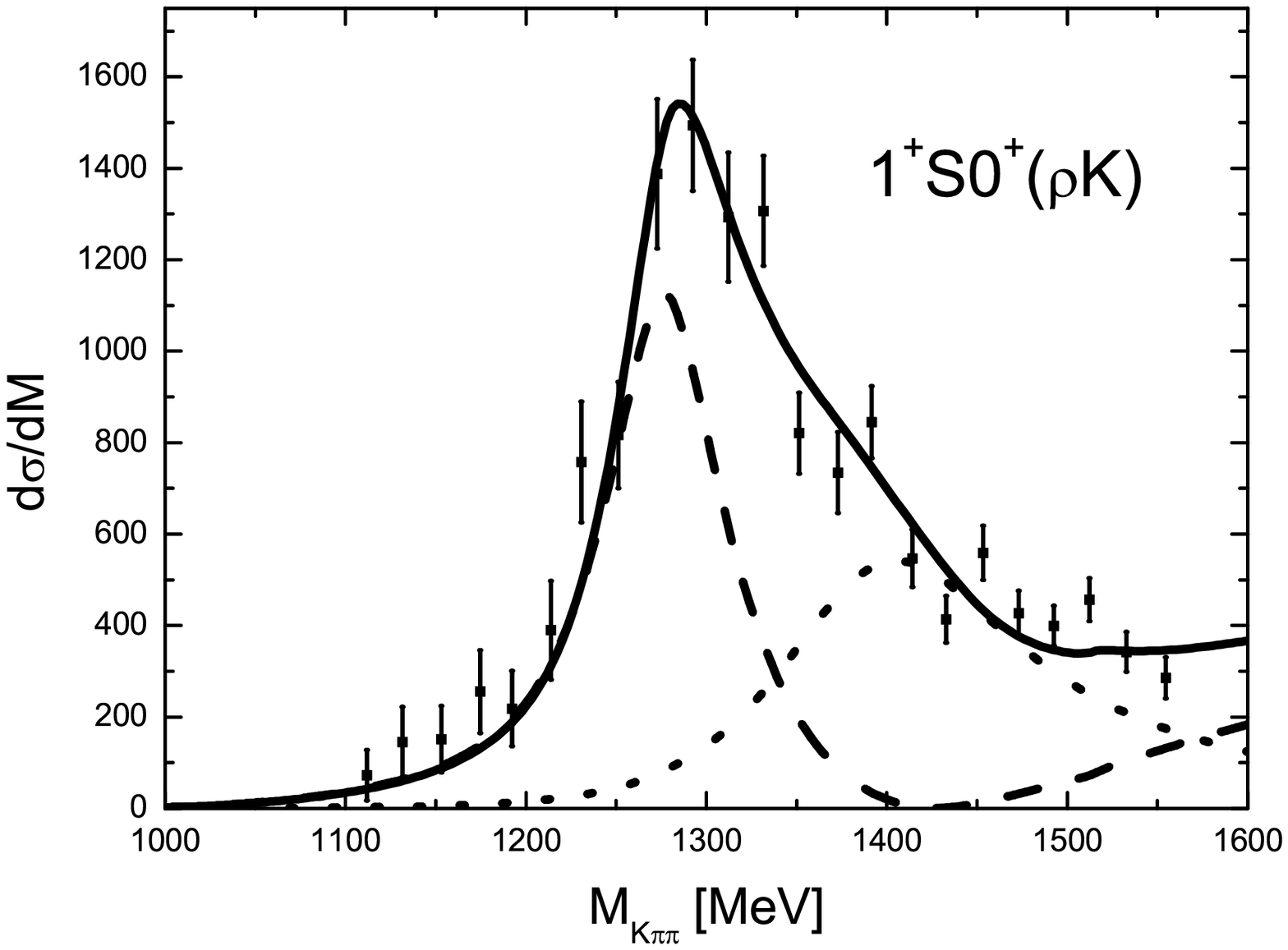}\\
\includegraphics[scale=0.4]{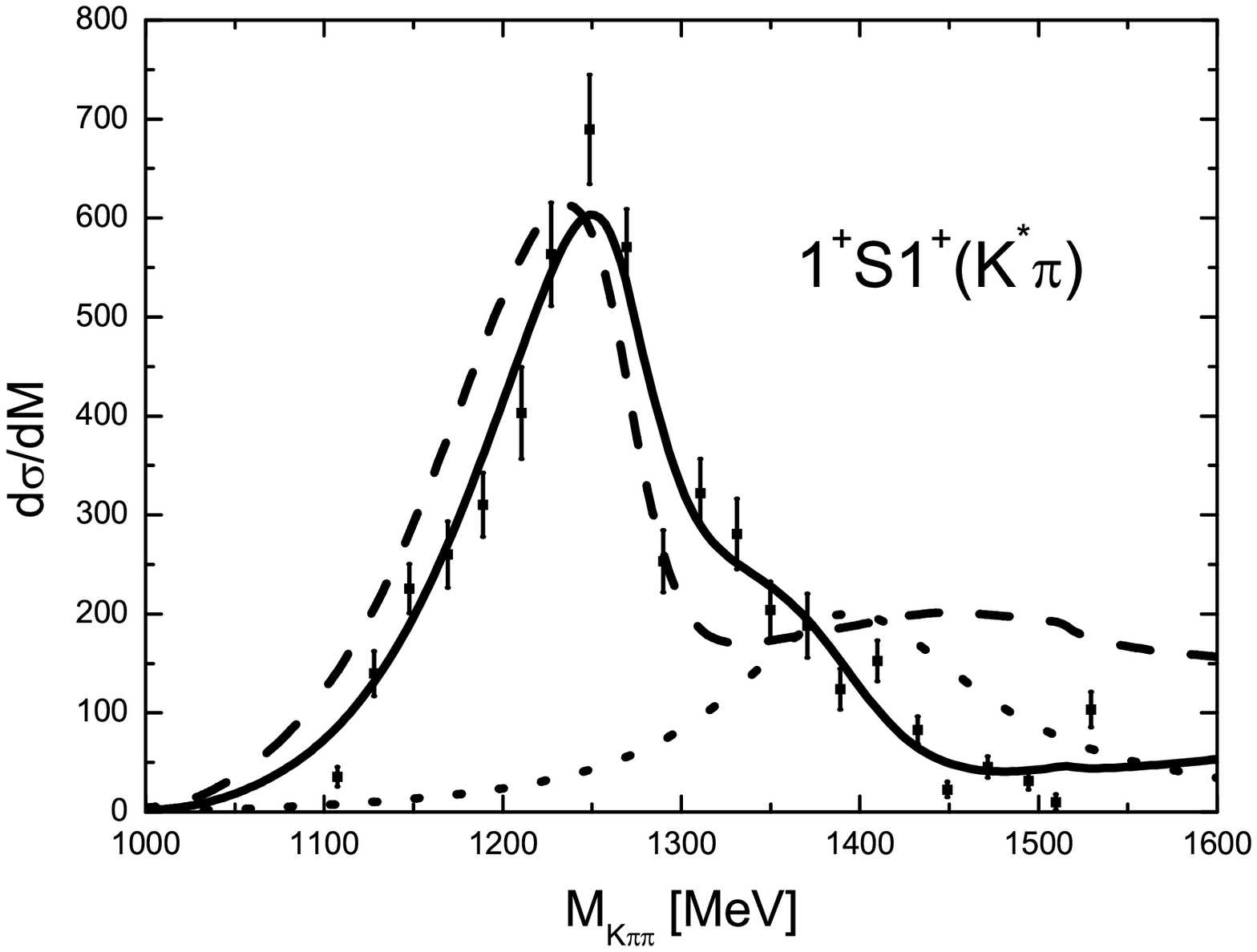}%
&\includegraphics[scale=0.4]{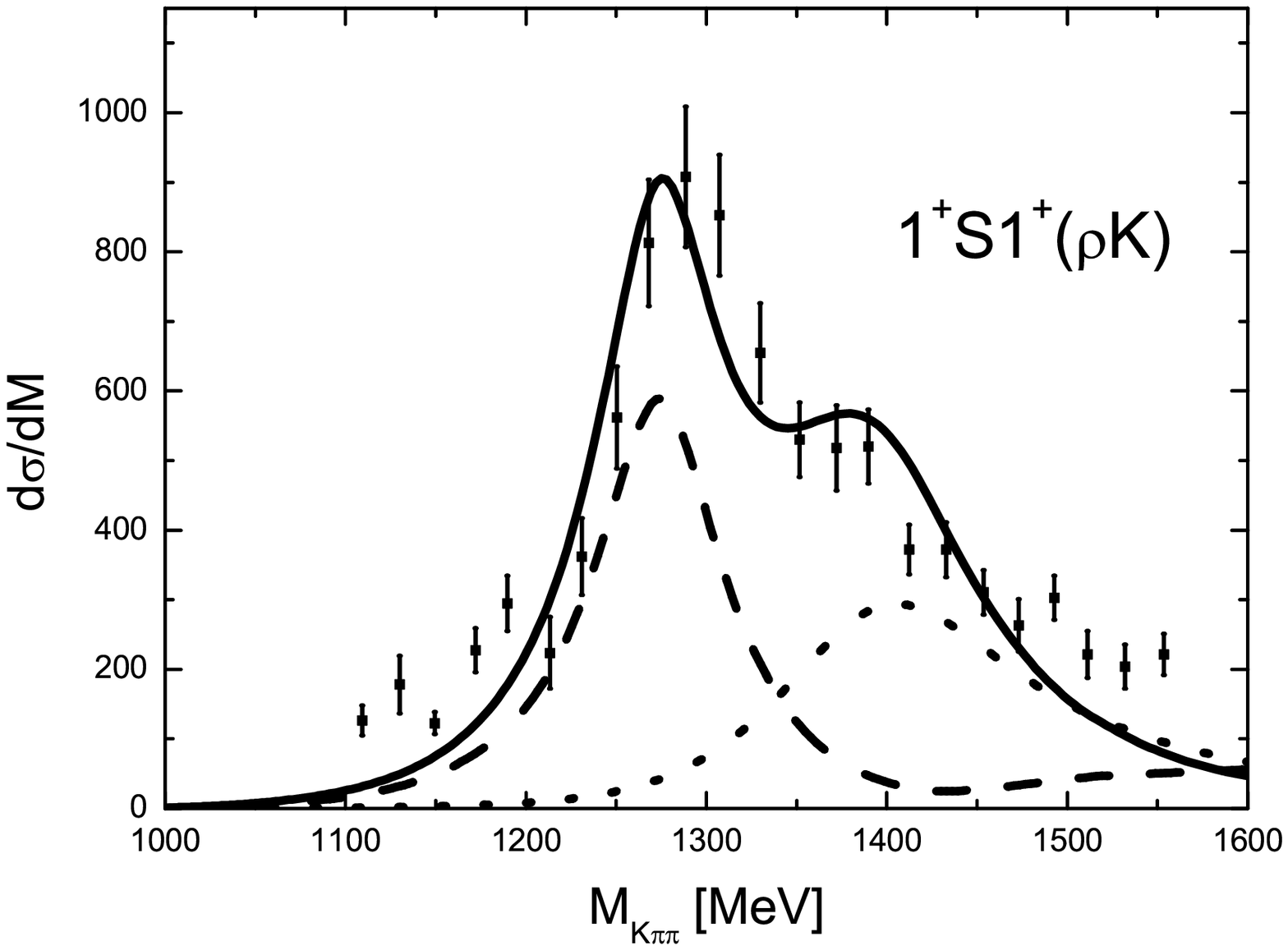}\\
\end{tabular}
\caption{The same as Fig.~3, but the theoretical fit is obtained by
including all the five channels: $\phi K$, $\omega K$, $\rho K$,
$K^*\eta$ and $K^*\pi$. }
\label{fig:CERN3}
\end{center}
\end{figure*}

Using these relations, we have refitted the WA3 data by assuming $D$
real, $F$ complex, $D'$ and $F'$ complex, given the arbitrariness of
a global phase. The results are shown in Fig.~\ref{fig:CERN3}.

It is seen that they in general resemble those obtained with two
channels.  The difference is mostly seen in the contribution of
$K_1(1400)$ in different channels. This can be easily understood
because the $K^*\eta$, $\omega K$, $\phi K$ channels start to
contribute at higher energies and thus distort the $K_1(1400)$
contribution seen in Fig.~\ref{fig:s1}.
 The above analysis showed that (i)
The most important channels are $K^*\pi$ and $\rho K$. (ii) The
SU(3) relations assume that the production vertices are due to pure
octet operators. Nevertheless, we find this assumption quite
reliable since the $K_1$'s resonances of the quark model are SU(3)
octets. (iii) The analysis must be understood as providing support
for our findings and conclusions with only the two channels
$K^*\pi$ and $\rho K$.

\section{Summary and conclusion}

In the present work, we have made a theoretical study of the
possible  experimental manifestation of the double pole structure of
the $K_1(1270)$,  predicted in a previous work using the techniques
of the chiral unitary approach to implement unitarity in the
vector-pseudoscalar meson interaction.  The model obtains two poles
in the $I=1/2$, $S=1$, vector-pseudoscalar scattering amplitudes
which can be assigned to two $K_1(1270)$ resonances. One pole is at
$\sim1200$\,MeV with a width of $\sim250$\,MeV and the other is at
$\sim1280$\,MeV with a width of $\sim150$\,MeV. The lower pole
couples more to the $K^*\pi$ channel and the higher pole couples
dominantly to the $\rho K$ channel. Different reaction mechanisms
may prefer different channels and thus this  explains the
different invariant mass distributions seen in various experiments.
We have analyzed the WA3 data on the $K^-p\to K^-\pi^+\pi^-p$
reaction since it is the most conclusive and high-statistics
experiment quoted in the PDG on the $K_1(1270)$ resonance. Our model
obtains a good description of the WA3 data both for the $K^*\pi$ and
$\rho K$ final state channels. In our model,  the peak in the
$K\pi\pi$ mass distribution around the $1270$\,MeV region is
a superposition of the two poles, but in the $K^*\pi$ channel the
lower pole dominates and in the $\rho K$ channel the higher pole
gives the biggest contribution.  It is worth stressing that the
physical properties quoted in the PDG obtained from the WA3 data
analysis rely, obviously, upon considering only one pole for the
$K_1(1270)$. These data have  long been interpreted by assuming
either a substantial Gaussian background or unitarized Deck
background. While the Gaussian background method is purely
empirical, we have shown that the Deck background method seems to
suffer from instability and critical destructive interferences. In contrast,
and as we mentioned above, the data can be explained in simpler terms by
our model, where every peak corresponds to resonances.
 Of course, this fact, although desirable, is not a physical requirement and more data 
should be analyzed to finally distinguish between our proposal, with two poles 
making up the $K_1(1270)$ resonance, and that with only one pole.

On the other hand, it is worth mentioning that in the literature,
 the $K_1(1270)$ and the $K_1(1400)$ resonances have always been
 considered as a mixture of the two corresponding SU(3)
 eigenstates. The mixing is defined by a mixing angle $\theta_K$. The
 value of this angle has been under extensive study for many years
 but no consensus has yet been reached. Recent studies seem to
 favor a value of $\sim 60\degr$ (see
 Refs.~\cite{Roca:2003uk,Li:2006we} for a short review). An
 interesting issue, however, has been raised by the BES study of
 charmonium decays to axial-vector plus pseudoscalar mesons. In
 $\psi(2s)$ decay, they found the $\psi(2s)\rightarrow
 K_1(1400)\bar{K} $ branching fraction is smaller than that for the
 $\psi(2s)\rightarrow K_1(1270)\bar{K} $ by at least a factor of 3.
 To accommodate this, one needs a mixing angle of $\theta_K<29\degr$.
 While in the $J/\psi$ decay, the $J/\psi\rightarrow
 K_1(1400)\bar{K} $ branching fraction is larger than the upper
 limit for the $J/\psi \rightarrow K_1(1270)\bar{K} $ mode. This
 would require a mixing angle of $\theta_K>48\degr$. As a possible future 
 application of our framework, these two different values for the mixing 
 angle could be possibly explained if one assumes that in
 the $\psi(2s)$ and $J/\psi$ decays, one actually sees the different
 two poles of the $K_1(1270)$ that we are referring in this work. 
 As the two poles would mix differently with the $K_1(1400)$, 
 they  would give rise to different values for the mixing angles. 
 In this sense, more high-statistics data of
 $\psi(2s)$ and $J/\psi$ decay would be very helpful to test the
 above assumption.

\section{Acknowledgments}
L. S. Geng acknowledges many useful discussions with Alberto
Mart\'{i}nez, Michael Doering, Kanchan Khemchandani and M. J. Vicente Vacas. We would
like to thank W. Ford for valuable comments regarding the
experiments used in this paper. This work is partly supported by
DGICYT Contract No. BFM2003-00856, FPA2004-03470, the Generalitat Valenciana, and the
E.U. FLAVIAnet network Contract No. HPRN-CT-2002-00311. This
research is part of the EU Integrated Infrastructure Initiative
Hadron Physics Project under Contract No. RII3-CT-2004-506078.


\begin{thebibliography}{99}

\bibitem{Yao:2006px}
  W.~M.~Yao {\it et al.}  [Particle Data Group],
  J.\ Phys.\ G {\bf 33}, 1 (2006).

%

\bibitem{Bai:1999mq}
  J.~Z.~Bai {\it et al.}  [BES Collaboration],
  Phys.\ Rev.\ Lett.\  {\bf 83}, 1918 (1999)
  [arXiv:hep-ex/9901022].

\bibitem{Roca:2005nm}
  L.~Roca, E.~Oset and J.~Singh,
  Phys.\ Rev.\ D {\bf 72}, 014002 (2005)
  [arXiv:hep-ph/0503273].

\bibitem{Magas:2005vu}
  V.~K.~Magas, E.~Oset and A.~Ramos,
  Phys.\ Rev.\ Lett.\  {\bf 95}, 052301 (2005)
  [arXiv:hep-ph/0503043].

  \bibitem{Armenteros64}
  R. Armenteros {\it et al.}
   Phys.\ Lett.\ B {\bf 9}, 207 (1964).

\bibitem{Astier:1969dt}
  A.~Astier {\it et al.},
  Nucl.\ Phys.\ B {\bf 10}, 65 (1969).

\bibitem{Firestone:1972st}
  A.~Firestone, G.~Goldhaber, D.~Lissauer and G.~H.~Trilling,
  Phys.\ Rev.\ D {\bf 5}, 505 (1972).

\bibitem{Gavillet:1978rj}
  P.~Gavillet {\it et al.}  [Amsterdam-CERN-Nijmegen-Oxford Collaboration],
  Phys.\ Lett.\ B {\bf 76}, 517 (1978).

\bibitem{Rodeback:1980zt}
  S.~Rodeback {\it et al.}  [CERN-College de France-Madrid-Stockholm
                  Collaboration],
  Z.\ Phys.\ C {\bf 9}, 9 (1981).

\bibitem{Brandenburg:1975gv}
  G.~W.~Brandenburg {\it et al.},
  Phys.\ Rev.\ Lett.\  {\bf 36}, 703 (1976).

\bibitem{Daum:1981hb}
  C.~Daum {\it et al.}  [ACCMOR Collaboration],
  Nucl.\ Phys.\ B {\bf 187}, 1 (1981).

\bibitem{Abe:2001wa}
  K.~Abe {\it et al.}  [Belle Collaboration],
  Phys.\ Rev.\ Lett.\  {\bf 87}, 161601 (2001)
  [arXiv:hep-ex/0105014].

\bibitem{Asner:2000nx}
  D.~M.~Asner {\it et al.}  [CLEO Collaboration],
  Phys.\ Rev.\ D {\bf 62}, 072006 (2000)
  [arXiv:hep-ex/0004002].

\bibitem{Bugg:2005ni}
  D.~V.~Bugg,
  Eur.\ Phys.\ J.\ A {\bf 25}, 107 (2005)
  [Erratum-ibid.\ A {\bf 26}, 151 (2005)]
  [arXiv:hep-ex/0510026].

\bibitem{Ablikim:2005ni}
  M.~Ablikim {\it et al.}  [BES Collaboration],
  Phys.\ Lett.\ B {\bf 633}, 681 (2006)
  [arXiv:hep-ex/0506055].

\bibitem{Otter:1976kk}
  G.~Otter {\it et al.}  [Aachen-Berlin-CERN-London-Vienna Collaboration],
  Nucl.\ Phys.\ B {\bf 106}, 77 (1976).

\bibitem{Vergeest:1979jd}
  J.~S.~M.~Vergeest {\it et al.}  [AMSTERDAM-CERN-NIJMEGEN-OXFORD
                  Collaboration],
  Nucl.\ Phys.\ B {\bf 158}, 265 (1979).

\bibitem{Bowler:1976qe}
  M.~G.~Bowler,
  J.\ Phys.\ G {\bf 3}, 775 (1977).

\bibitem{Weinberg:1978kz}
  S.~Weinberg,
  PhysicaA {\bf 96}, 327 (1979).

\bibitem{Gasser:1984gg}
  J.~Gasser and H.~Leutwyler,
  Nucl.\ Phys.\ B {\bf 250}, 465 (1985).

\bibitem{Meissner:1993ah}
  U.~G.~Meissner,
  Rept.\ Prog.\ Phys.\  {\bf 56}, 903 (1993)
  [arXiv:hep-ph/9302247].

\bibitem{Bernard:1995dp}
  V.~Bernard, N.~Kaiser and U.~G.~Meissner,
  Int.\ J.\ Mod.\ Phys.\ E {\bf 4}, 193 (1995)
  [arXiv:hep-ph/9501384].

\bibitem{Pich:1995bw}
  A.~Pich,
  Rept.\ Prog.\ Phys.\  {\bf 58}, 563 (1995)
  [arXiv:hep-ph/9502366].

\bibitem{Ecker:1994gg}
  G.~Ecker,
  Prog.\ Part.\ Nucl.\ Phys.\  {\bf 35}, 1 (1995)
  [arXiv:hep-ph/9501357].

\bibitem{Lutz:2003fm}
  M.~F.~M.~Lutz and E.~E.~Kolomeitsev,
  Nucl.\ Phys.\ A {\bf 730}, 392 (2004)
  [arXiv:nucl-th/0307039].

\bibitem{Dobado:1996ps}
  A.~Dobado and J.~R.~Pelaez,
  Phys.\ Rev.\ D {\bf 56}, 3057 (1997)
  [arXiv:hep-ph/9604416].

\bibitem{Oller:1997ng}
  J.~A.~Oller, E.~Oset and J.~R.~Pelaez,
  Phys.\ Rev.\ Lett.\  {\bf 80}, 3452 (1998)
  [arXiv:hep-ph/9803242].

\bibitem{Oller:1998hw}
  J.~A.~Oller, E.~Oset and J.~R.~Pelaez,
  Phys.\ Rev.\ D {\bf 59}, 074001 (1999)
  [Erratum-ibid.\ D {\bf 60}, 099906 (1999)]
  [arXiv:hep-ph/9804209].

\bibitem{Oller:1998zr}
  J.~A.~Oller and E.~Oset,
  Phys.\ Rev.\ D {\bf 60}, 074023 (1999)
  [arXiv:hep-ph/9809337].

\bibitem{Oller:1997ti}
  J.~A.~Oller and E.~Oset,
  Nucl.\ Phys.\ A {\bf 620}, 438 (1997)
  [Erratum-ibid.\ A {\bf 652}, 407 (1999)]
  [arXiv:hep-ph/9702314].

\bibitem{Birse:1996hd}
  M.~C.~Birse,
  Z.\ Phys.\ A {\bf 355}, 231 (1996)
  [arXiv:hep-ph/9603251].

\bibitem{Flatte:1976xu}
  S.~M.~Flatte,
  Phys.\ Lett.\ B {\bf 63}, 224 (1976).

\bibitem{Carnegie:1976cs}
  R.~K.~Carnegie, R.~J.~Cashmore, M.~Davier, W.~M.~Dunwoodie, T.~A.~Lasinski, D.~W.~G.~Leith and S.~H.~Williams,
  Nucl.\ Phys.\ B {\bf 127}, 509 (1977).

\bibitem{Daum:1980ay}
  C.~Daum {\it et al.}  [ACCMOR Collaboration],
  Nucl.\ Phys.\ B {\bf 182}, 269 (1981).

\bibitem{Bowler:1974th}
  M.~G.~Bowler, J.~B.~Dainton, A.~Kaddoura and I.~J.~R.~Aitchison,
  Nucl.\ Phys.\ B {\bf 74}, 493 (1974).

\bibitem{Bowler:1975my}
  M.~G.~Bowler, M.~A.~V.~Game, I.~J.~R.~Aitchison and J.~B.~Dainton,
  Nucl.\ Phys.\ B {\bf 97}, 227 (1975).

  \bibitem{Georgi} H. Georgi, {\it Lie Algebras in Particle Physics},  
 (Westview Press, Colorado, 1999), 2nd ed.

\bibitem{Roca:2003uk}
  L.~Roca, J.~E.~Palomar and E.~Oset,
  Phys.\ Rev.\ D {\bf 70}, 094006 (2004)
  [arXiv:hep-ph/0306188].

\bibitem{Li:2006we}
  D.~M.~Li and Z.~Li,
  Eur.\ Phys.\ J.\ A {\bf 28}, 369 (2006)
  [arXiv:hep-ph/0606297].

\end{thebibliography}
\end{document}